\documentclass[letter,12pt]{article}
\usepackage{graphicx,amssymb,amsmath,epstopdf,color,hyperref}

\def\beq{\begin{equation}}
\def\eeq{\end{equation}}
\def\bea{\begin{eqnarray}}
\def\eea{\end{eqnarray}}

\def\gev{\, {\rm GeV}}
\def\tev{\, {\rm TeV}}
\newcommand{\gsim}{\lower.7ex\hbox{$\;\stackrel{\textstyle>}{\sim}\;$}}
\newcommand{\lsim}{\lower.7ex\hbox{$\;\stackrel{\textstyle<}{\sim}\;$}}

\def\xsection#1{\section{#1}}

\begin{document}

\begin{flushright}
April 4, 2019
\end{flushright}

\vspace{0.07in}

\noindent
\begin{center}

{\bf\large Discovery goals and opportunities in  high energy physics: \\ a defense of BSM-oriented exploration over signalism}

\vspace{1.3cm}
{ James D. Wells}\footnote{email: jwells@umich.edu}

{\it Leinweber Center for Theoretical Physics \\
University of Michigan, Ann Arbor, MI 48109 USA} \\

\end{center}

\vspace{2cm}

\noindent
{\it Abstract:} 

Discoveries come through exclusions, confirmations or revolutionary findings with respect to a theory canon populated by the Standard Model (SM) and beyond the SM (BSM) theories.  Guaranteed discoveries are accomplished only through pursuit of BSM exclusion/confirmation, and thus require investment in the continual formation and analysis of a vibrant theory canon combined with investment in experiment with demonstrated capacity to make BSM exclusions/confirmations.   Risks develop when steering away from BSM-oriented work toward its methodological rival, ``signalism," which seeks to realize SM falsification or revolutionary discoveries outside the context of any BSM rationale. It is argued that such an approach leads to 
inscrutable  exertions that reduce prospects for all  discovery.  The concepts are applied to the European Strategy Update, which seeks to identify future investments in forefront experiment that bring a balance of guaranteed and prospective value.

\vfill\eject

\tableofcontents

\vfill\eject

\vfill\eject
\xsection{Introduction}

The practice of science includes a wide range of activities, ranging from theoretical speculations  to experimental analysis. These activities are all in the pursuit of scientific discovery ---  securely knowing something of science value that we did not know before. In this essay a formalization of the language of discovery is put forward that articulates common ambient notions in high-energy physics. From this, an argument is made that persistent and guaranteed discovery, as well as  enhanced prospects for discovery of every kind,  are accomplished through the co-work of beyond the Standard Model (BSM) theory and experimental work focused on BSM exclusion/confirmation. Signalism is the main methodological rival to BSM-centered exploration. It proposes to achieve SM falsification or revolutionary discoveries without any reference to BSM theories. However, it will be argued that signalism is an inscrutable and non-rational methodology for science discovery and puts at risk all types of discovery as conventionally conceived.

The thesis introduced above should not be interpreted to imply lower value or  lesser status for other activities such as SM theory work, SM experimental analysis, formal theory, or detector/experimentation development. These, as we shall see, are indispensable activities ultimately in the service of discovery when done well. Nevertheless, it is argued that positioning BSM as the central attractor of theory work and experiment discovery is what guarantees, vindicates,  and gives meaning to those other efforts. 

This essay is admittedly long. The impatient reader can go straight to the summary (sec.\ 10) to read a listing of the main points developed. The full essay aims to give context, justification and nuance to those claims. Secs.\ 2-5 set up the conceptualization of discovery, with arguments and illustrations for the BSM-centered approach peppered throughout. Sec.\ 6 addresses the methodological rival ``signalism" more directly, and suggests that it comes up short compared to BSM-centered work. In some sense, sec.\ 6 is the culmination of the main thesis of the essay that BSM-centered work is superior to signalistic approaches for the pursuit of discovery. Secs. 7 and 8 illustrate the main points of the essay through discussion of recent discoveries of gravity waves and the Higgs boson, and also through discussion of the European strategy update, which aims to make possible more discoveries in the future.  Sec.\ 9 discusses the risks and signs of discovery ending, and their antidotes. Sec.\ 10 summarizes the essay.

\subsection{Theoretical vs.\ experimental discovery}

Let us continue the introduction by first discussing a little more on what is meant by ``discovery" in this essay. 
Colloquially we refer to discoveries mainly within the experimental realm. There are exceptions, such as speaking of Einstein having discovered General Relativity, whereas Eddington only confirmed it experimentally, or rather discovered a unique predicted feature of the theory (bending of light).  However, the majority of cases where the appellation discovery is applied is reserved to experimental work: Thomson discovered the electron;  Rutherford discovered the proton; Chadwick discovered the neutron;  Anderson and Neddermeyer discovered the muon; Richter's and Ting's collaborations discovered the $J/\psi$; the Gargamelle collaboration discovered neutral currents; the CDF and D0 collaborations of Fermilab discovered the top quark; Atlas and CMS collaborations of CERN discovered the Higgs boson; etc.  

Standard usage of discovery in science rightly puts the primary emphasis on experiment. Applying the word ``discovery" for the invention of a theory, whenever it does happen, as in the case of General Relativity, often only takes place after experimental confirmation, which is the strongest form of discovery. Discovery has the sense of uncovering something that is true that was lying in wait for us to find. For us,  theories will not be evaluated in our forthcoming discussion on whether they always existed or whether they are permanently  true fixtures waiting to be found, but rather whether they are presently adequate in the face of all experimental results known. Thus, it would be preferable perhaps to replace phrases such as ``she discovered the theory of X" with something less provocative, such as  ``she educed the theory of X". 

Colloquially we may also utilize the word ``discovery" for the product of a ``founder of discursivity," as a Foucauldian might say, where a new work produces ``the possibilities and the rules for the formation of other texts"~\cite{Foucault:1984}, where ``other texts" in our context are forthcoming scientific works made possible by the founding work. A key example of this in recent years was the ``discovery" of warped extra dimensions by Randall and Sundrum~\cite{Randall:1999ee}, which resulting in a multitude of additional works that built upon their founding idea. 
When Randall and Sundrum were appropriately awarded the 2019 Sakurai Prize\footnote{The Sakurai Prize is the highest award given by the American Physical Society for work in theoretical particle physics.} their citation was for ``in particular the discovery that warped extra dimensions of space can solve the hierarchy puzzle..."~\cite{Sakurai:2019}.
The word ``discovery" is implicitly modified by ``theoretical" by the context of the award being exclusively in the theoretical domain. However, there has been no experimental verification (not yet at least)  of warped extra dimensions. Therefore, by the common implicit rules of scientific discourse one could not say in a contextless environment that ``warped extra dimensions have been discovered." Only after experimental verification could one presume to make such a grand statement. For this reason, the unmodified word ``discovery" in a contextless sentence must necessarily refer to a result confirmed by experiment, such as ``the discovery of general relativity", or ``the discovery of neutrinos," etc.

Nevertheless, theory plays a significant role in the discovery process. Many times experimental discoveries are made because they constructed dedicated apparatuses to search in subtle places that theory suggested. The most celebrated recent example of that is the discovery of the Higgs boson, which required a multi-billion dollar experiment with special particle detectors designed primarily with the Higgs boson discovery requirements in mind. Thus, any full accounting of discovery must also make theory an integral part of the story.

\subsection{Experiment as transformations of the theory canon}

The key construct through which we account for theory's role in discovery is what can be called the {\it theory canon}. The theory canon is the collection of all theories devised, including the standard reference theory (i.e., the Standard Model in particle physics), that satisfy all the requirements that physicists believe make these theories good descriptions of nature. There are many such requirements. Some are uncontroversial (i.e., must satisfy all known experimental data, must be mathematically consistent), while others are controversial (i.e., must be natural, must be simple, must not be in swampland). It is not just theorists who decide what belongs in the canon, but all stakeholders that test such theories. For this reason what is admitted into the theory canon is a difficult community discussion. 

More will be said about the theory canon later, but let us suppose we have one. Experimental discoveries are then made within the context of that canon. Confirmations are made when a theory or a key component of a theory within the canon is confirmed. Exclusions are made with respect to a theory in the canon. (One cannot exclude what one does not know.) Similarly, relegation or falsification of a theory to the dustbin of history (i.e., total exclusion) is an experimental discovery that can only be achieved if there is a theory canon within which the falsified theory had once lived. The existence of the theory canon enriches experiment and makes possible numerous discoveries that were otherwise inconceivable.

Of course, there are experimental discoveries that take place completely outside the context of the theory canon. Finding completely unexpected particles or interactions or signals that are unanticipated by any theory within the theory canon is revolutionary. Such revolutionary discoveries (e.g., discovery of the muon is thought to be one such discovery) are part of physics history and presumably should continue to be into the future. Nevertheless, it should be noted that what makes them spectacular, eye-popping, revolutionary and rare is the existence of an advanced theory canon that is exploded by the discovery.

In the following, three broad categories of experimental discovery are described: confirmation, exclusion, and revolutionary. There are important further distinctions and subcategories that will be made within these broad categories, which includes SM confirmations, BSM confirmations, falsifications of the SM, falsification of a BSM theory, or falsification of the entire theory canon. As stated at the top, it will be argued that perpetual and guaranteed discovery passes through the focusing gateway of BSM theory and BSM-centered experiment.

\subsection{The work of assured discovery}

It is hoped that articulating the concepts, categories and paths of discovery will contribute to assessing valuable activity in high-energy physics enterprise, especially as we plan for its future. As we contemplate all the aspects of guaranteed discovery, we see that the effort that gives rise to it can be organized into three core discovery activities that must be healthy for high-energy physics to be healthy: 
\begin{itemize}
\item ``model building": constructing a vibrant and motivated BSM theory canon.
\item ``theory analyzing": connecting theory canon ideas with phenomenological implications.
\item ``experimental work": translating phenomenological implications of the theory canon into experiments with assured confirmation/exclusion capability. 
\end{itemize}
All three of these are necessary, and require intense, focused and unique knowledge and skill sets. 

The categories above are based on action-oriented work, not static labels of individuals, since a scientist can in principle participate in any combination of these three activities, although he/she most often has hard-won primary expertise in only one. Often a physicist can have substantial overlap in nearest-neighbor activities. For example, a physicist can contribute  to ``model building" and ``theory analyzing" and yet another can contribute to ``theory analyzing" and ``experimental work". 
 It is hoped that it will become clear after the argument is presented that if any of these activities dwindles, guaranteed discovery ends. 

The above list may give the reader the wrong impression that more formal theoretical work is viewed here as less relevant to discovery and less important.  Formal work includes many areas of active research including  string theory, AdS/CFT theory, amplitudes theory, finite temperature field theory, black hole conundrumology, information theory, etc.  Although formal work  looks far removed from discovery it is recognized by most to contribute as feed-in fuel  to  model building and theory analysis. For example, the proof of renormalizability of the weak interactions was critical to progress in concretizing the SM into a fully calculable theory (see Veltman's and `t Hooft's essays in~\cite{Hoddeson:1997}). As another example, the work of dualities in supersymmetric Yang-Mills theory~\cite{Seiberg:1994pq} led to significant developments in BSM model building~\cite{Csaki:1996zb}. Likewise, AdS/CFT correspondence~\cite{Maldacena:1997re} has given much deeper and fruitful correspondences between theories of warped extra dimensions and walking technicolor~\cite{ArkaniHamed:2000ds}, which on the surface looked unconnected. The recent development in the theory of amplitudes (for reviews, see, e.g.,~\cite{Elvang:2013cua,Cheung:2017pzi}) is hoped to one day provide a significantly better approach to theory analysis, and perhaps even model building. Similarly, in the past, the mathematical physics work of group theory, topology, differential geometry, etc.\ also could not have been spoken of directly as ``model building" or ``theory analyzing" as discussed above, yet they ultimately have played central roles in both. 

One could then interpret formal work as vital ``pre-discovery" work in the service of model building and theory analysis which is, in turn, in the service of (experimental) discovery. It is no less an important activity as any of the others for a healthy and vibrant field that wishes to continue making discoveries far into the future. 
Nevertheless,  if a particular activity of formal physics cannot be plausibly argued to have some possible connection to the  three more direct discovery activities (model building, theory analyzing, experimental work), then it is at risk of being a less relevant activity. 
It is a subtle task to evaluate formal theory work's ultimate relevance to discovery. That topic will be taken up elsewhere. For the purposes of this essay we need merely acknowledge that the ``pre-discovery" work of formal physics is crucial and contributes fuel to sustained progress in model building and theory analysis.

Lastly, just as formal work within theory gives fuel to future advances in model building and theory analyzing, so does ``pre-discovery" work in experimental physics. Detector R\&D, accelerator physics research, computational and electronics hardware advances, and analysis software tools, all contribute toward and seed progress in experimental work. In some sense this is the experimental analogue to theory's ``formal work," which is less direct and proximate to actual discoveries, but is vital work that enables more direct discovery activities to realize themselves consistently into the future.

\xsection{The theory canon}

One of the aims of theoretical physics is to seek  theories that have predictive capacity and are empirically adequate. Empirical adequacy is the ability of at least one point of the theory's parameter space to match all experimental measurements  simultaneously within a stipulated domain of applicability. If there exists one point in parameter space that is empirically adequate there usually exist an infinite number of point that are empirically adequate -- a ``good" region of parameter space. For example, in the SM there are an infinite number of input parameter points that match the data, such as the infinite number of top quark mass values in the experimentally allowed range $172.26\pm 0.61\gev$~\cite{Sirunyan:2018mlv}.

There can be numerous theories that are consistent with all known data. For example, in addition to the SM there is the minimal supersymmetric standard model~\cite{Martin:1997ns}, the next-to-minimal supersymmetric standard model, the minimal composite Higgs theory~\cite{Panico:2016}, the $SU(2)$ left-right gauge theory model, the minimal warped extra dimension model~\cite{Randall:1999ee}, the large extra dimension model~\cite{ArkaniHamed:1998rs}, the SM Effective Theory (SMEFT) theory with higher dimensional operators~\cite{Buchmuller:1985jz,Brivio:2017vri}, etc. All of these are called beyond the Standard Model (BSM) theories. Each of these theories has its own parameter space and may have different extended domains of applicability\footnote{By extended domains of applicability it is meant that a theory may purport to have a definite range of validity, such as a minimal supersymmetric theory up to the grand unification scale. Or, it may have augmented purposes compared to the SM, such as providing a dark matter candidate. This is the case of new theory that looks like the SM except it has, for example, one more real scalar $S$ that couples to the Higgs boson and is postulated to be the dark matter of the universe~\cite{Davoudiasl:2004be}.} beyond the minimal domain required for the SM success.

The collection of all theories that are empirically adequate are candidates for admission into the theory canon. Certainly the SM is within the theory canon, since it is the agreed-upon standard reference theory that agrees with the data.   In addition the SM, the theory canon contains the union of every empirically adequate BSM theory that a non-trivial subset of the expert scholarly community deems to have value beyond the SM. In addition to the SMEFT, the manifesting practice is to admit new theories into the canon that ``explain more" than the SM, such as dark matter, fermion mass and mixing angle hierarchies, small Higgs mass, small cosmological constant, coexistence with gravity, origins of spacetime symmetries and structure, origins of internal symmetries, coexistence with unification,  baryon asymmetry of the universe, etc. 

It must be repeatedly emphasized that every theory within the theory canon must be at present consistent with all known experiment. Furthermore, every theory within the canon is there provisionally and is never safe. Additional theory analysis can find that a theory is incompatible with an experiment in a way that had not before been understood. Or, an experiment can release a new experimental result that ejects theories from the canon. The parameter spaces of theories within the canon are continually under revision. The theory canon  shape-shifts often based on theoretical and experimental progress. Occasionally it is even annihilated when new revolutionary experimental results are inconsistent with every theory within the canon. In that case, a new theory canon is reborn, perhaps slowly, in the wake of this revolutionary development.

A key concept put forward in this work is that experimental discovery can and should be understood and categorized in terms of the effect it has on the theory canon. The above discussion of the theory canon and  its relevance to experiment is somewhat abstract and general by constructive necessity, but the implications are very concrete and the recognitions of various types of discovery are straightforward. In the following sections various forms of discovery are defined with respect to the theory canon, with specific examples provided to give greater clarity and applicability to the abstract notions. But first, we must say a word more about the standard reference theory (i.e., the SM of particle physics) and how it is represented in the full theory canon space.

As mentioned above, the categories of discovery that are discussed in more detail are delineated by the action that experiment does on the theory canon. It is useful to have a visualization of these various actions. In order to do that we must develop a visual representation of the theory canon based on the principles discussed earlier of what is in the canon. 

\begin{figure}[t] 
\begin{center}
\includegraphics[width=0.47\textwidth]{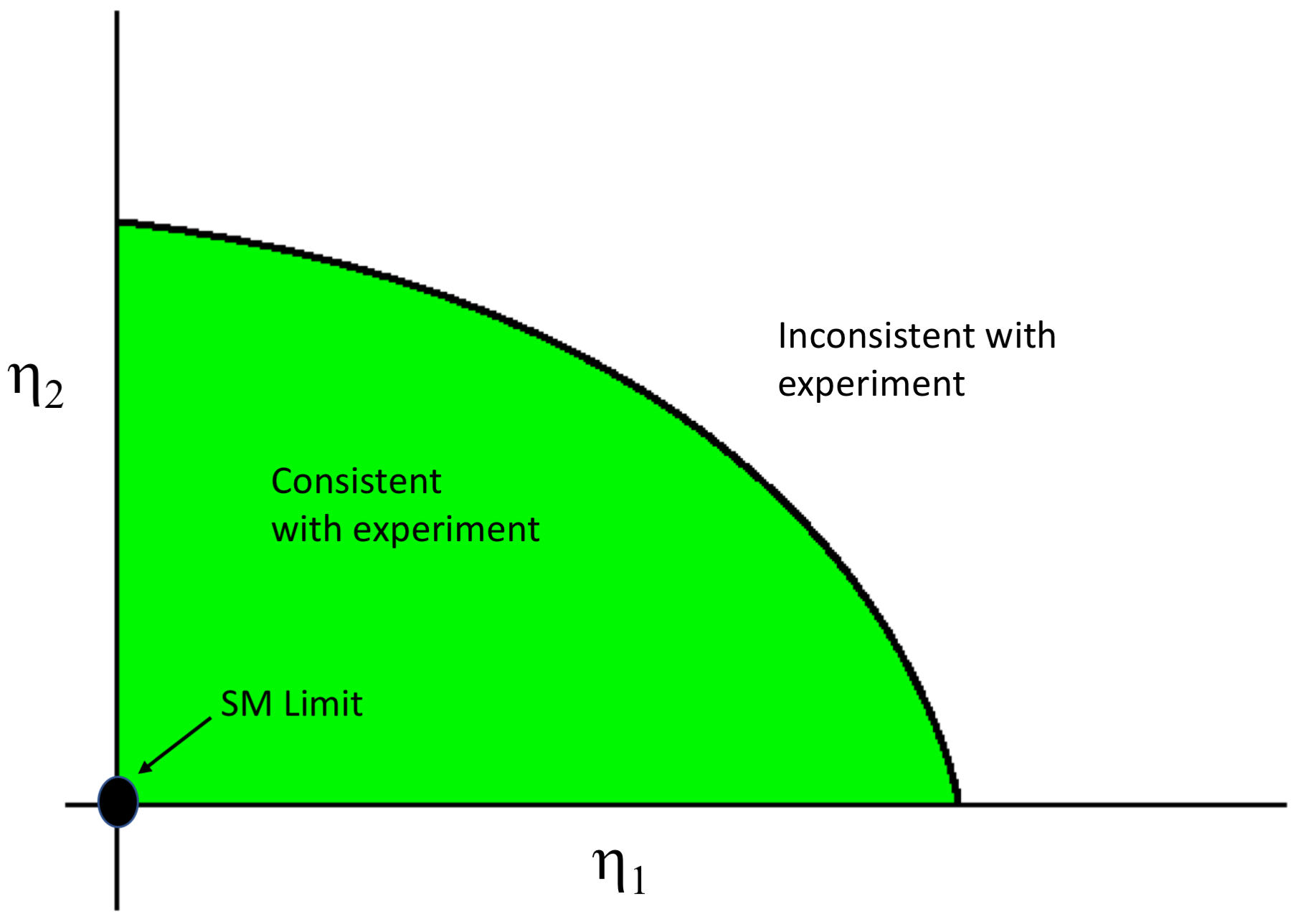} 
\caption{Green is the currently allowed parameter space of a BSM theory that decouples to predictions for observables indistinguishable from those of the SM when $\eta_{1,2}\to 0$.}
\label{fig:Slide1}
\end{center}
\end{figure}

We can visualize the theory canon as a collection of all admitted BSM theories, each of which is represented by a parameter space where the allowed region is demarcated (for us, in green), as seen in Fig.~\ref{fig:Slide1}. Here, the BSM theory has parameters $\eta_1$ and $\eta_2$ such that when $\eta_i\to 0$ all observables reproduce SM values. Thus, $\eta_{1,2}$ are decoupling parameters of a decoupling BSM theory, which is currently the most representative type of BSM theory within the theory canon.  The BSM theory may have many more parameters than just these two graphically shown, but the concept is the same: The SM is represented as a limiting point at the origin. Non-decoupling theories may not have a SM point anywhere in the visualization, but for it to be in the theory canon it must have experimentally allowed regions of parameter space. An important feature of a non-decoupling theory is that it can be ruled out even if the SM is exact in nature.

To take advantage of this decoupling behavior in visualizing a specific BSM theories, it is helpful to recast the BSM parameters such that the SM decoupling limit is always at the origin. In other words, instead of plotting $m_{1/2}$ vs.\ $m_0$, where the SM limit is really the ``point at infinity", we construct the inverse as parameters, e.g., $\eta_{1/2}=m_Z/m_{1/2}$ and $\eta_0=m_Z/m_0$, where the origin of the $(\eta_{1/2},\eta_0)$ is the decoupling limit of the SM\footnote{Minimal supersymmetry is not exactly a decoupling theory in the sense that the Higgs mass is computable in terms of superpartner masses and is not a free to be any value in the low-scale SM effective theory. For this reason, the allowed parameter space will never include exactly the origin in the $(\xi_0,\xi_{1/2})$ parameter space, or equivalently at the point  at infinity in the $(m_0,m_{1/2})$ parameter space, as illustrated for example by the allowed region of Fig.~1 of~\cite{Bagnaschi:2018igf} being restricted to finite values in the $(m_0,m_{1/2})$ plane.}. One can then represent the allowed region more compactly, and scientific progress and discovery is a tighter push  the origin with confirmation discovery potential ever present.

Another example is ``sequential  hypercharge $Z'$" theory. This is a new $Z'$ boson that couples to the SM in exactly the same way as the hypercharge gauge boson except that its mass $M_{Z'}$ and overall coupling strength $g_{Z'}$ are free parameters. This is arguably not within the theory canon since its motivation may not be high enough, but it demonstrates visualization in an especially simple way, which is analogous to many theories that are within the theory canon, such as dark photon dark matter theories. The $\xi$-variables for this representation are $g_{Z'}$ and $m_Z/M_{Z'}$ which gives decoupling (i.e., SM predictions) at the origin. It is also true that the entire $g_{Z'}=0$ and $m_Z/M_{Z'}=0$ axes are in the decoupling limit as well. It is this reason that experimental constraints on the parameter space will push closer and closer to the $g_{Z'}=0$ and $m_Z/m_{Z'}=0$ axes but can never get there. As the exclusion capability increases  it nevertheless does open the opportunity for a signal to develop in that previously unexplored region of parameter space.
That would constitute an important discovery.

As alluded to above, there may be other theories that have no decoupling limit at all to the SM. For example, the case of minimal no-scale supergravity theories with neutralino dark matter LSP do not allow superpartner masses to decouple~\cite{Diehl:1994ff}.
This theory is similar to the standard minimal constrained supersymmetric standard model except that $m_0=0$ is required, which puts an upper bound on $m_{1/2}$, otherwise the LSP is no longer a neutralino and so cannot be the dark matter. The upper bound on $m_{1/2}$, and thus lower bound in $\xi_{1/2}=m_Z/m_{1/2}$, prevents reaching a decoupling limit within the theory. 

One might object that minimal no-scale supergravity is just a subset of the parameter space within the more expansive minimal supergravity theories that do not require $m_0=0$ and thus should not be consider as an additional theory within the theory canon.  However, landmarks of experimental progress are powerfully stated as total exclusion of coherent, self-contained BSM theories with specifically motivated theoretical structures and phenomenological targets (such as dark matter, $g-2$ explanation, etc.). Recognitions of BSM falsifications are powerful milestones, which can in turn also impact  views of the community on the larger category of theories, such as whether minimal supersymmetry should still occupy high table in the theory canon. 

We have discussed minimal supersymmetry and minimal $Z'$ models within this discussion of the theory canon. But there are many more ideas of high interest to the high-energy physics community including warped extra dimensions, twin Higgs theories, little Higgs theories, minimal scalar-extended dark matter scenarios, superlight vector dark matter, low-scale baryogenesis sector theories, etc. The stature of various theories within the theory canon is not the subject of this essay, yet it must be recognized that various ideas are promoted and others relegated  as their strengths and weakness are revealed in the intense theoretical and experimental scrutiny they experience. In this sense there is value in some ``group-think" activity to promote, criticize and explore ideas. A thousands scientists in a thousand attics working on a thousand totally distinct ideas are unlikely to make the progress needed for discovery. Likewise, a thousand scientists in one attic working on only one idea is also unlikely to engender a healthy flow of ideas and discovery. As with most such endeavors, a balance between these two extremes toward constructing and analyzing the theory canon is likely to most useful. However, balance in this sense is not to be recommended to exist within every individual, but rather across the field, since individuals must focus to make impact. Partly for this reason, banishing an idea from the theory canon that has many invested proponents is not easy. Nevertheless, theory ideas die regularly, albeit it quietly with few visiting the graves (minimal technicolor, minimal non-supersymmetric $SU(5)$ GUTs, supersymmetric electroweak light-stop baryogenesis, minimal conformal SM, etc.). 

Let us also remark that it is entirely reasonable to be cautious of theory talk about such lofty aims as ``elucidating the true structure of space and time" and ``constructing deeper reformulations of the laws of nature," etc. A new, improved language is not particularly transformative if one cannot order a good dinner with it, as every speaker of Esperanto can attest.   Less controversial is a more instrumentalist appraisal of knowledge gain and theory development, which assesses the ability to predict that ``if I do A, then I know B will come next", where, of course, B can be a collection of probabilistic outcomes. This power of prediction is worth more than any fancy subtle theory or ``deep insight" into the soul of nature\footnote{Distinguishing true science from mere visionary pronouncements has been a difficult problem for millinia. Nevertheless, as scholars frequently note, ``we have come to realize that the best proof that our knowledge is genuine is that it enables us to do something"~\cite{Farrington:1969}.}. However --- and this can never be forgotten --- powerful workhorse predictive theories are often given birth by lofty theory/mathematical parents (e.g., non-abelian gauge symmetries, general coordinate invariance, supersymmetric theories, conformal theories, etc.). Thus, erring on the side of inclusive acceptance to theory development is in order, but researchers in theoretical high-energy physics should be able to articulate how their work is (or at least ``might be") connected to the construction of new BSM theories that answer outstanding problems in nature (i.e., ability to make predictions or to explain ``histories"), or they should be able to explain how their work enables (or at least ``might enable") more effective analysis of the SM and BSM canon theories that enlarges capacity for exclusion/confirmation discoveries. Theory work that can do neither is unlikely to contribute to genuine discovery.

Finally, our purpose here is not to develop an evaluative theory of what should and should not be in the theory canon, or a praxis theory of how some theories get promoted and others banished among empirically adequate alternatives, or other such philosophical concerns. The purpose here is mainly to point out that a theory canon does indeed exist, as any high-energy physicist recognizes. They are the theories that many people continue to work on. They are the theories that experimentalists aim to find or constrain. They are the theories that end up in technical design reports motivating new experiments. Furthermore, the theory canon exists even though individual physicists might differ on what the community views as being contained within it, especially some theories on the ``edges" of the canon (somewhat fewer practitioners, less experimental interest, remaining allowed parameter space is extremely ``small" compared to prior motivated assessments, etc.). Criticisms, promotions, additions and deletions of the theory canon will always be a part of high-energy physics. Nevertheless, discovery is and should be made with respect to that canon, as will be developed more fully below. These discoveries are confirmation, exclusion and revolutionary, to which we now turn.

\xsection{Confirmation discoveries}

With respect to the theory canon, there are three kinds of confirmation discoveries. The first kind of confirmation, SM feature confirmation, is experimentally verifying a feature of the SM that hitherto had not yet been established, or even was viewed by many as highly uncertain. The second type of confirmation, SM locus confirmation, is confirming by experiment the empirical adequacy of a narrowly carved locus of points in SM parameter space motivated by additional principles that go beyond the SM definition (i.e., BSM motivated).  And a third type of confirmation discovery, BSM confirmation, is
verifying a feature of a BSM theory by which the SM is eliminated from the theory canon and the BSM theory is elevated to the new SM.

\subsection{SM feature confirmation}

Let us first consider a SM confirmation. Throughout the history of particle physics there are many such examples. Notable ones in recent years include discoveries of the charm quark~\cite{Augustin:1974xw,Aubert:1974js}, of the $W$ boson in 1983~\cite{Arnison:1983rp,Banner:1983jy}, the top quark in 1995~\cite{Abe:1995hr,D0:1995jca}, and the Higgs boson in 2012~\cite{Aad:2012tfa,Chatrchyan:2012xdj}.  The charm quark and Higgs boson discoveries were particularly momentous since confidence that they should be found was not uniform among high-energy physicists. In addition to finding evidence for these elementary particles, a  SM confirmation discovery can be said to involve any qualitative property or manifestation of the theory that had not yet been observed. Examples of these include the presence of CP violation in B decays~\cite{Abe:2001xe,Aubert:2001nu}, discovered in 2001, observation of CP violation in charm decays~\cite{Aaij:2019a}, discovered very recently, and the existence of three active species of light neutrinos~\cite{Blondel:2016mrw}. 

The determination of three neutrino species was partially achieved before the start of LEP's $Z$-pole experiments in 1989, where a review then noted  that data was consistent with $N_\nu=2.0^{+0.6}_{-0.4}$, and concluded that ``$N_\nu=3$ is perfectly compatible with all data. Although the consistency is significantly worse,  four families still provide a reasonable fit. In the framework of the Standard Model, a fifth light neutrino is, however, unlikely"\cite{Denegri:1989if}. 
Very quickly after the turn-on of LEP and SLC, measurements of invisible final states of the $Z$ width suggested $3.12\pm 0.19$ as of October 1989~\cite{Blondel:2016mrw}. By the time a final analyses were being completed on the precision electroweak data at LEP/SLC the precision completely ruled out anything but $3$ neutrino species ``assuming that only invisible $Z$ decays are to neutrinos coupled according to SM expectations"~\cite{ALEPH:2005ab}. In other words, the SM feature of three neutrinos had been confirmed.

Every SM  confirmation discovery is momentous since it both signifies a leap in experimental sophistication and it secures knowledge that we could not be sure of before. In addition, it expels speculations (i.e., BSM theories) that certain features of the SM are indeed absent or altered in the true underlying theory. For example,  disbelief and dramatic alternatives to the SM Higgs explanation of electroweak symmetry breaking and mass generation thrived within the theory canon up to the moment of the Higgs discovery~\cite{Wells:2018nwj}, underscoring the importance of SM confirmation discoveries. 

One of the subtleties about a SM confirmation discovery resides in the meaning of confirmation. Confirmation colloquially often implies that one thing (theory, fact, etc.) was found to be true beyond any doubt and all relevant alternatives are not (``I have confirmed that Thurston attended the opera last Wednesday night.") This is too restrictive of a notion for confirmation in scientific research that aims to go beyond ``for all practical purposes" for those satisfied with the needs of the here and now and who feel no compelling desire to delve deeper.  However, our task is to continually refine our explanations for physical phenomena. Confirmation for us then is achieving a strong localization within the theory canon, without requiring that the localization returns one and only one empirically adequate theory. 

This subtlety regarding confirmation reared its head in the early days of the Higgs discovery when CERN scientists were hesitating calling what they found a ``Higgs boson." Instead, they came up with other phrases such as ``new particle consistent with the Higgs boson"~\cite{CERN:2012a,Aad:2012tfa} and ``new particle ... with spin different than one"~\cite{Chatrchyan:2012xdj}. The reason for this is that there were initially many other theories in the canon (dilaton scalars, spin-2 resonances, etc.) that could have explained the measurements they were getting at those early stages. As the CMS collaboration concluded, ``The results presented here are consistent, within
uncertainties, with expectations for a standard model Higgs boson. The collection of further
data will enable a more rigorous test of this conclusion"~\cite{Chatrchyan:2012xdj}.
 As the data accrued and some of the more exotic ideas (e.g., heavy graviton-like objects) were becoming more inconsistent with the measurements, CERN scientists felt more and more comfortable in 2014, nearly two years after its first discovery, to simply declare that ``it has been identified as a Higgs boson"~\cite{Chatrchyan:2014vua}.

Nevertheless, precisely what does it mean to say the Higgs boson has been discovered? If it means a scalar boson that has all the decay branching partial widths of the textbook SM Higgs boson field to six significant digits, then nobody can say we have confirmed that. Experimental uncertainty combined with the existence of many ideas within the theory canon that can give small deviations within experimental allowances forbid us from declaring with  certainty that what we label as the Higgs boson in the SM is indeed what has been discovered\footnote{Indeed, several future colliders, such as ILC~\cite{Moortgat-Picka:2015yla}, HL-LHC and HE-LHC~\cite{Cepeda:2019klc} and CLIC~\cite{Abramowicz:2016zbo} are being proposed to discover BSM theories that give altered Higgs boson phenomena in subtle ways~\cite{Curtin:2013fra}.}. Instead, more precisely we have discovered a narrowed localization in the theory canon consistent with the existence of a scalar boson and consistent with all the properties of the SM Higgs boson to within measurement errors. Some properties have not been measured to within even an order of magnitude of its predictions (e.g., triple Higgs coupling) whereas other properties have been measured to within about 15\% (e.g., Higgs decays to photons and $Z$ bosons)~\cite{Atlas:Higgs,CMS:Higgs}. 

Despite the caveats, one does now hear said that ``the Higgs boson has been confirmed," with the implication that this SM feature has been confirmed. This simple proclamation is acceptable since as a community we know that it is short-hand for ``events have been registered in the detector that are consistent with what would be created by the existence of a SM Higgs boson, and whose precision measurements are sufficient to highly suggest that indeed something rather close the SM Higgs boson was found if not the SM Higgs boson itself, although measurements in even the near term might force us to relabel the object again as a Higgs-like particle that shares properties with the SM Higgs but does not quite have exactly SM Higgs properties due to small deviations measured in its couplings to other states compared to those derived from precision SM analysis." Such implicit Joycean statements are the bane of all confirmation discoveries, but they do reveal more accurately the nature of confirmations, whose exasperating tentativeness rewards us nevertheless with the seeds of future possible discovery.

\subsection{SM locus confirmation}
\label{sec:SMlocusconf}

The parameters of the standard theory are never measured perfectly. For example, a parameter may be known to within a factor of two, and later measured to within 1\% after dedicated experimental study.  Whenever there are uncertainties in the parameters there remains the prospect of hypothesizing a higher structure on the standard theory that predicts what those parameters converge to when a much better measurement is made later. Or there might be a relation between parameters that is required by assuming an additional symmetry structure on top of the minimal symmetries required to define the standard theory. A ``SM locus confirmation discovery" is when, upon further experimental improvements, a BSM theory-derived locus of points in the SM parameter space is measured to indeed be the experimentally selected region. 

\begin{figure}[t] 
\begin{center}
\includegraphics[width=0.47\textwidth]{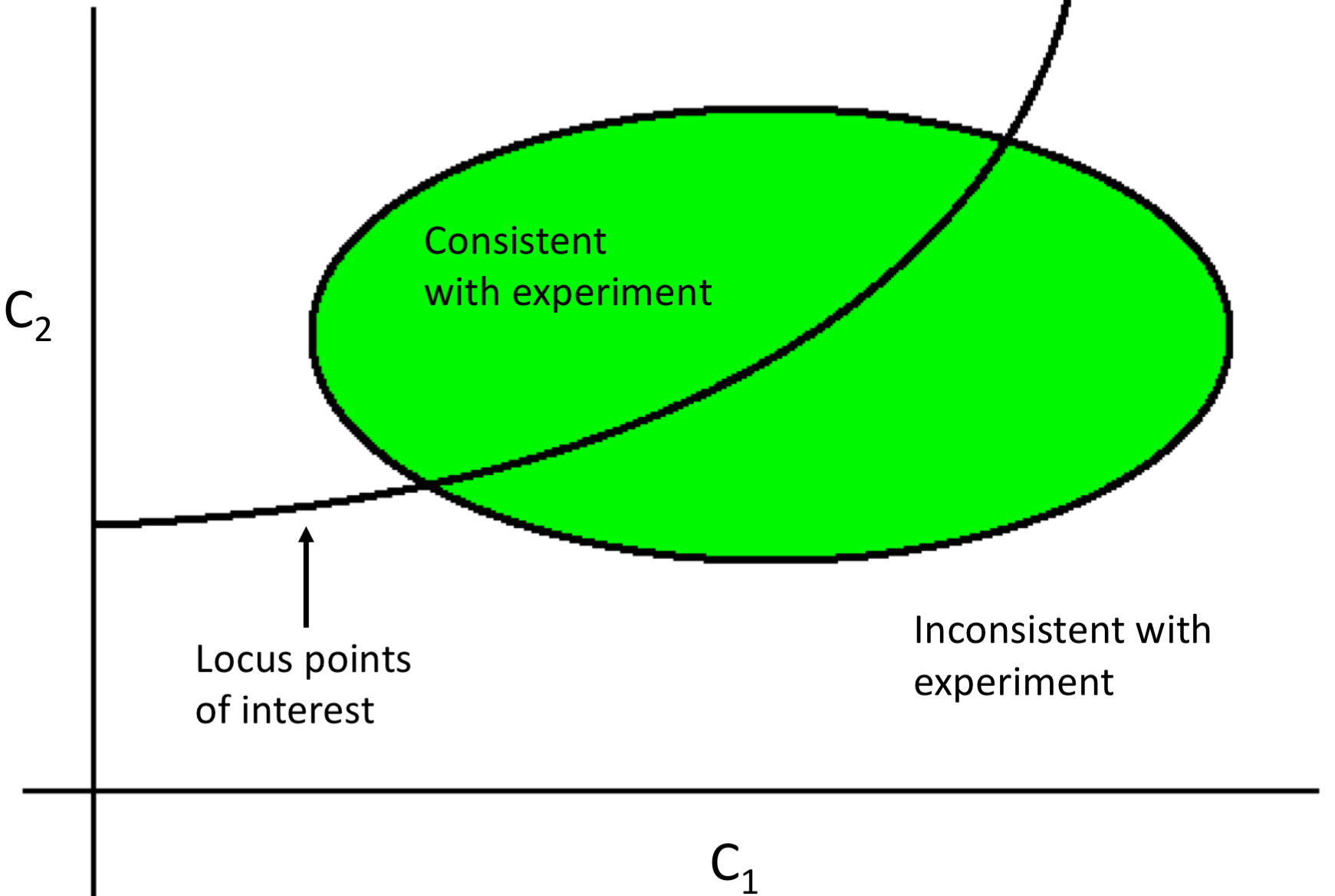} 
\caption{The green region shows the  allowed region by all current experiment  of two SM parameters $C_1$ and $C_2$. The solid black line is a ``locus of interest" from the point of view of BSM theory that reduces to the SM in the low-energy limit. This BSM prediction can then be tested by future experiment.}
\label{fig:Slide4}
\end{center}
\end{figure}

\begin{figure}[t] 
\begin{center}
\includegraphics[width=0.47\textwidth]{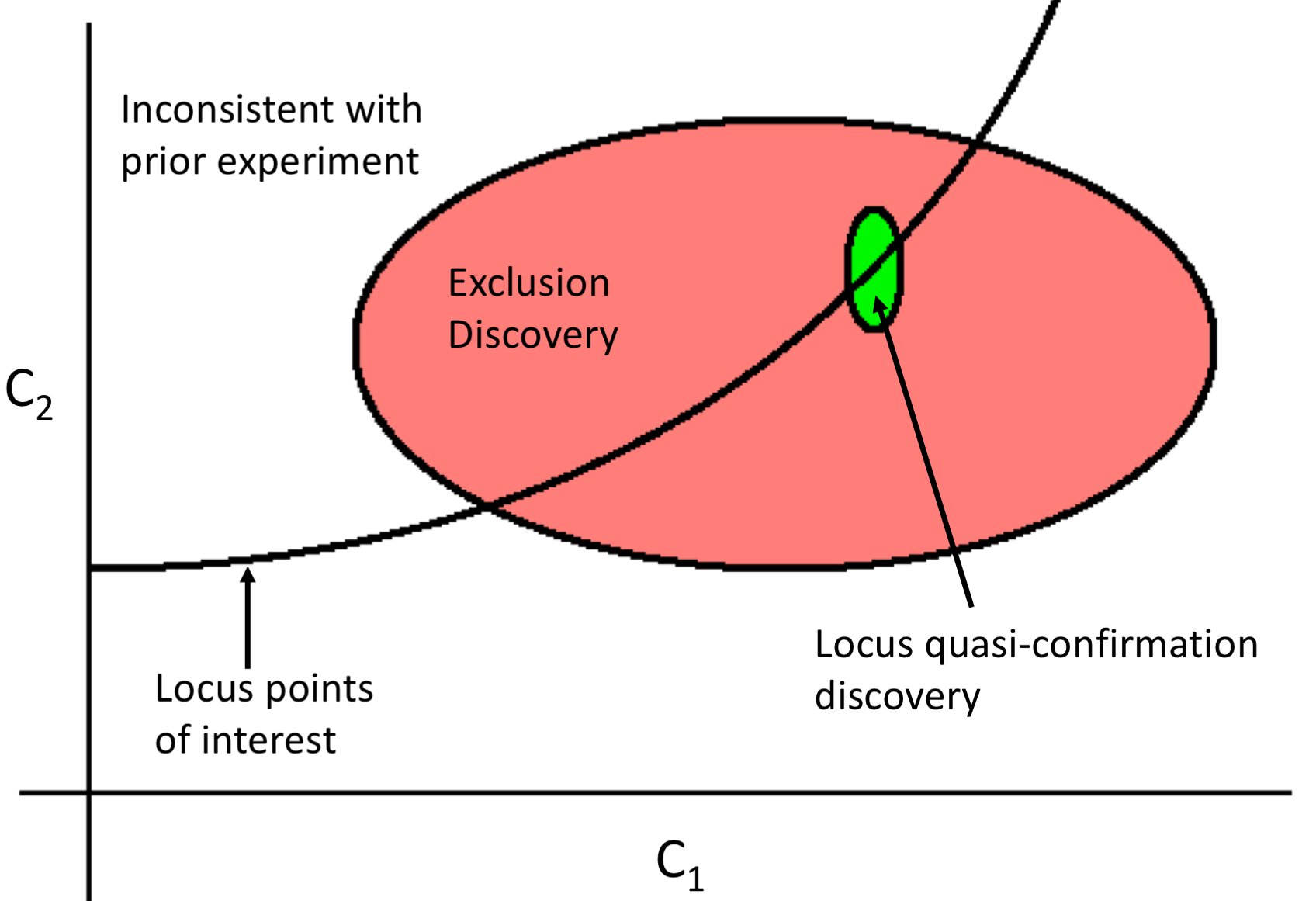} 
\caption{After further experiment, much of the $(C_1,C_2)$ parameter space of the SM is excluded except for a small remaining green region that is centered on the ``locus points of interest" predicted by a BSM theory. This is only a quasi-confirmation discovery since there are points allowed by experiment that nature could select that are off the locus points of interest line.}
\label{fig:Slide5}
\end{center}
\end{figure}

One way that a locus prediction arises is when analyzing a motivated BSM theory that is in full flourish only at higher energy scales where the new particles and forces have their characteristic scales. Upon integrating out this BSM theory and proceeding to a lower energy SM effective theory, the constraints of the full theory may lead to a tight restriction on what values the SM parameters may take. A graphical illustration of this is given in Fig.~\ref{fig:Slide4} where an experimentally allowed region in the parameter space of two SM variables $c_1$ and $c_2$ is shown in green. The SM treats $c_1$ and $c_2$ as independent variables, but the BSM theory predicts that the relation between $c_1$ and $c_2$ is fixed and given by the black curved line in the figure. If future experiment makes significant progress and new measurements localize at the locus points of interest, this is a locus confirmation discovery. Fig.~\ref{fig:Slide5} schematically depicts this type of discovery. 

The challenge with a locus confirmation discovery such as that depicted schematically in Fig.~\ref{fig:Slide5} is that the confirmation might not survive additional theory or experiment scrutiny. On the theory side, it is always the case that calculations might not have been complete or correct, and the locus of points identified were incorrectly positioned. Such errors are simply errors and must be rectified, just as there is the possibility of experiment making error. More subtle is when to apply the label ``confirmation" even if every thing were done correctly by theorists and experimentalists. In Fig.~\ref{fig:Slide5} the black line of locus points is thinner than the extent of the new green experimentally allowed region. Although the experimental focussing on this a priori established locus of points is very impressive, it is also possible that future experiment will result in a significantly smaller green allowed region that does not overlap with the locus of points of interest on the black line. In such a case, the notion of ``confirmation" would have to be retracted. For this reason, one may wish to call the discovery depicted in Fig.~\ref{fig:Slide5} a ``quasi-confirmation discovery"  rather than a confirmation discovery, since it is not guaranteed by any means that the discovery will hold up after further experimental results.

\begin{figure}[t] 
\begin{center}
\includegraphics[width=0.47\textwidth]{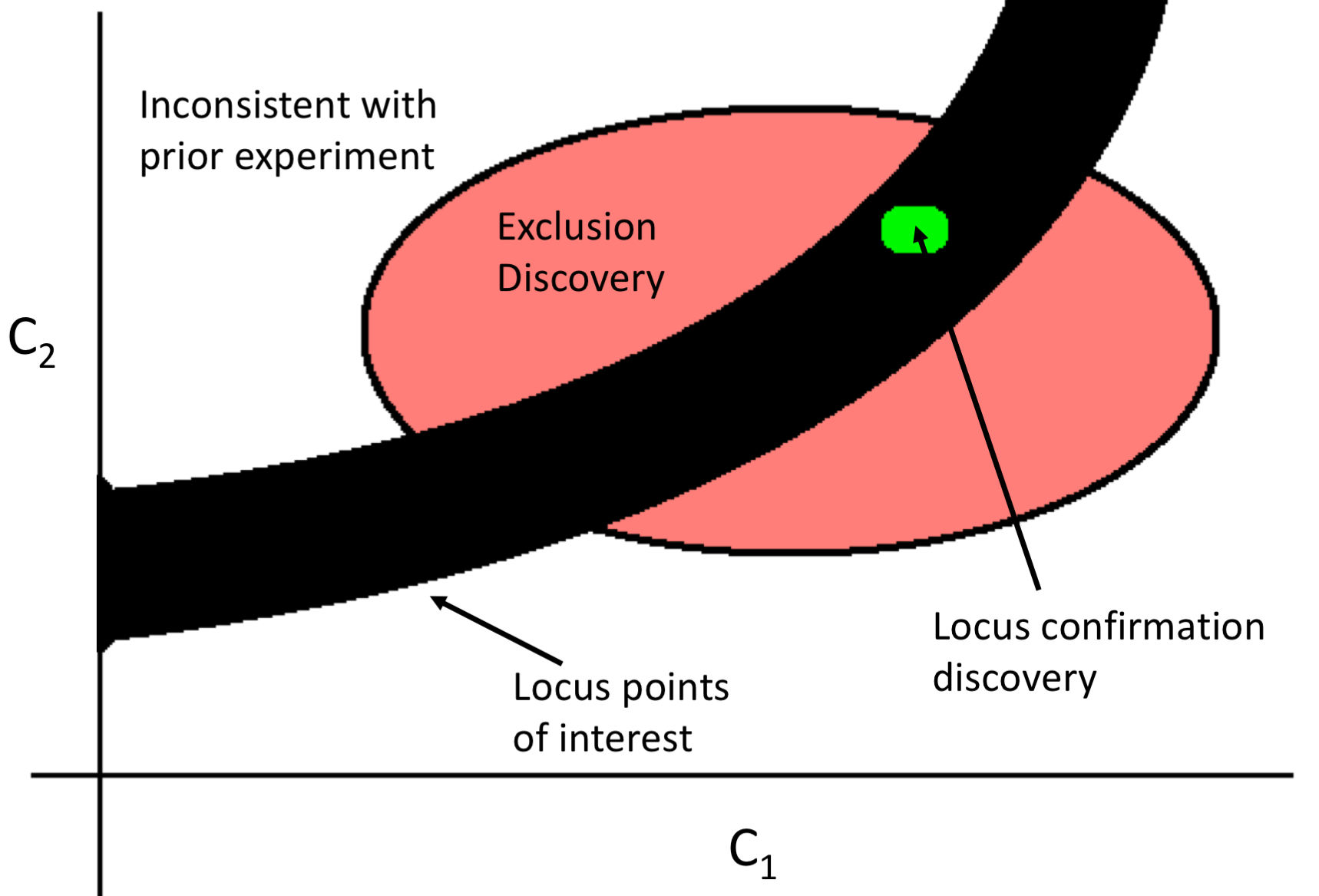} 
\caption{After further experiment, much of the $(C_1,C_2)$ parameter space of the SM is excluded except for a small remaining green region that is centered on the ``locus points of interest" predicted by a BSM theory. This is a locus confirmation discovery since the experimentally allowed region is entirely within the locus points of interest.}
\label{fig:Slide7}
\end{center}
\end{figure}

On the other hand, if the transformative experiment results in a new green region of experimentally points that is fully contained within a continuous locus of points of interest, that indeed would be a true locus confirmation discovery. Such a discovery would not be subject to new categorization from new experiments in the future, unless of course experiments had made mistakes of a systematic nature. This kind of discovery is depicted in Fig.~\ref{fig:Slide7}.

Locus confirmation discovery does not mean that the BSM theory that gave rise to that locus has been ``confirmed" or ``discovered." Only the locus has been confirmed. This can create increasing interest in the BSM theory and lead to investigations on how other more direct tests, involving qualitatively new phenomena, may be devised that could lead to a more direct BSM confirmation discovery, as will be discussed in sec.~\ref{sec:BSMconfirmation}. 

There are numerous instances of locus confirmation discoveries in recent high energy physics. Minimal supersymmetry with supersymmetry scale less than a TeV 
predicted that the Higgs boson mass should be less than about $135\gev$  before the Higgs boson was found\footnote{See Martin's discussion~\cite{Martin:1997ns} on p.54 of version 1 from 1997 which put the upper limit on MSSM light CP-even Higgs mass at $\lsim 130\gev$ and then on p.95 of  version 4 from 2011 (just prior to Higgs boson discovery), which put the upper limit at $\lsim 135\gev$ from improved supersymmetric Higgs mass calculations.}.  In this case, the locus of points of interest was all masses below $135\gev$ for the lightest scalar Higgs bosons. The discovery of the existence of the Higgs boson, as predicted and in opposition to other Higgsless theories, and second that its mass was $125\gev$, thus less than $135\gev$, is considered by many an important success of the theory. Nevertheless, although it is an interesting locus confirmation discovery, it is obviously not to be considered a supersymmetry confirmation discovery, as we emphasized in more general terms above.

\subsection{BSM confirmation}
\label{sec:BSMconfirmation}

The third  kind of confirmation discovery, BSM confirmation, occurs when experiment excludes the SM (the reference theory) and localizes around the parameter space of a BSM theory. Many BSM theories could be consistent with the new-found localization within the theory canon, and so many BSM theories could rightly lay claim to a confirmation discovery. What is key is that the SM is excluded and at least one BSM theory in the theory canon remains empirically adequate.

To schematically represent a BSM confirmation discovery, we revert back to our BSM parameterization illustration of two variables $\eta_1$ and $\eta_2$ such that when $\eta_i\to 0$ all observables reproduce SM values.  In Fig.~\ref{fig:Slide1} that we introduced earlier, we saw a large green region where the BSM theory is perfectly consistent with all known data within the target observables of the theory\footnote{A ``target observable of a theory" is an observable that the theory is designed to compute and purports to be correct.}, and beyond the green region the theory is inconsistent with data for any number of reasons. Perhaps there is an additional state that should have been seen by the Tevatron, or perhaps the theory points there are inconsistent precision $Z$ decay observables from LEP measurements, etc. 

\begin{figure}[t] 
\begin{center}
\includegraphics[width=0.47\textwidth]{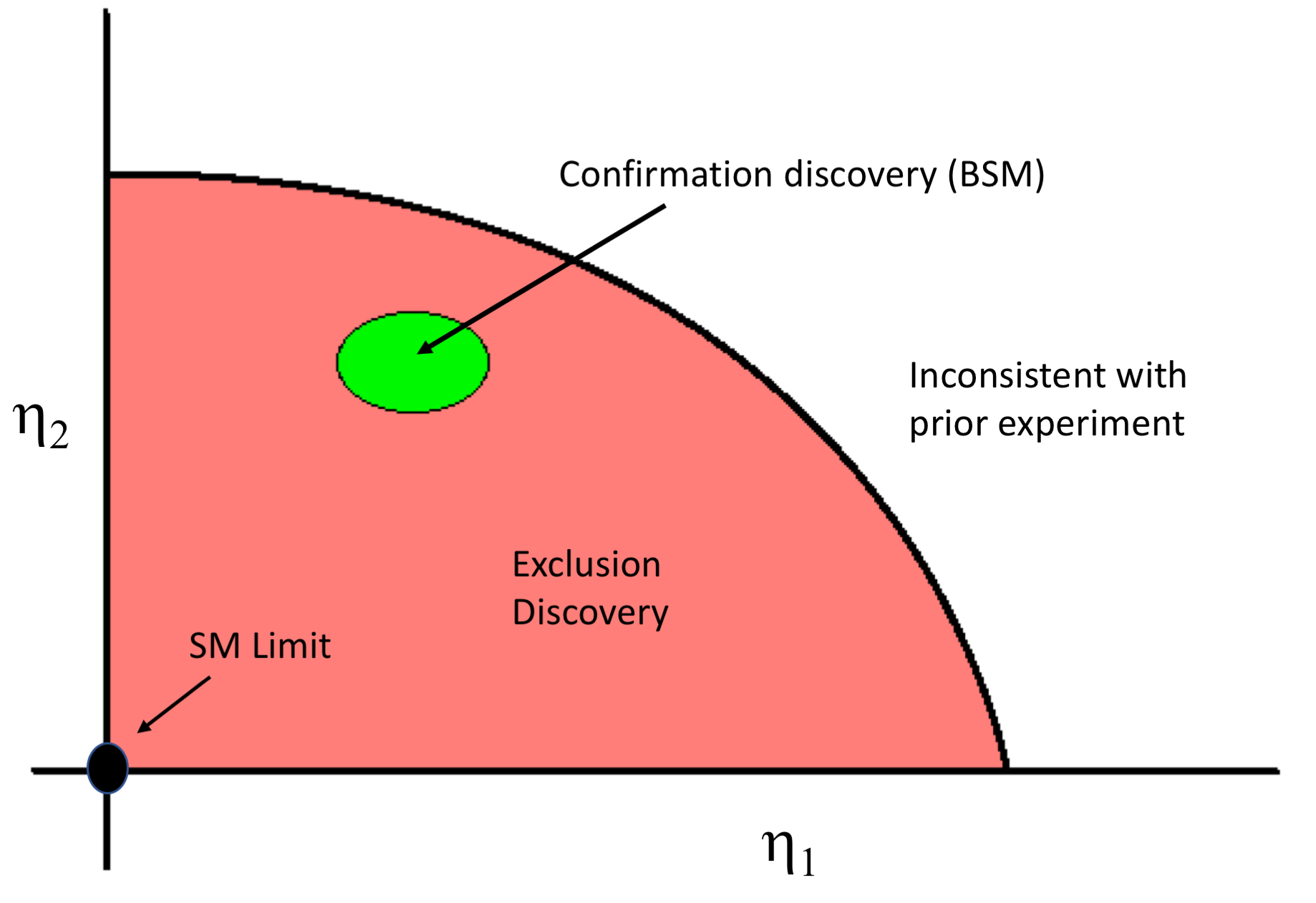} 
\caption{In a BSM theory that includes parameters $\eta_1$ and $\eta_2$, and where SM predictions are obtained in the limit of $\eta_{1,2}\to 0$, one obtains a BSM confirmation discovery if future experiment rules out the origin and converges on an allowed region (green region) where $\eta_{1,2}\neq 0$. }
\label{fig:Slide3}
\end{center}
\end{figure}

Now, a new experiment runs that has discovery potential.  In other words, the new experiment can either confirm or exclude green regions of $\eta_1-\eta_2$ parameter space in Fig.~\ref{fig:Slide1} after its run. This would be a transformative experiment, in the sense that parameter space that we thought before was viable is either confirmed or excluded by virtue of the experiment. Let us now suppose that after this new transformative experiment has run its course, the only points in the $\eta_1-\eta_2$ plane that are consistent with the data are those that do not include the origin. In other words, experiment has shown that the SM is inconsistent with the data, while at the same time the BSM theory under consideration is consistent with experiment. In that case an enclosed green region of the  $(\eta_1,\eta_2)$ plane is selected, as shown in  Fig.~\ref{fig:Slide3}. Such an outcome would signify a BSM confirmation discovery. Note, by definition, SM falsification is a necessary byproduct of any BSM confirmation, where such a definition has the helpful additional implication that it prevents too eager researchers from conflating SM locus confirmation with BSM confirmation.

A BSM discovery would not mean we have necessarily found the unique correct theory of nature, just as prior to the transformative experiment we could not say that the SM was the uniquely correct theory. Indeed, the existence of any two or more  theories that are consistent with the data is proof enough against the notion of a uniquely correct theory. Nor can we say that the uniquely true and correct theory underneath everything must be one of the ones that we have already contemplated (i.e., currently in the theory canon). In fact, it is a defensible conjecture that no theory can be complete and inviolable that emerges from finitely equipped minds and survives finitely scoped experiment, which are the twin rickety foundations for all theories. 

There are numerous examples of BSM confirmation discoveries in the history of physics. One of the most celebrated early such discovery was the positron (anti-electron) posited by Dirac in 1931~\cite{Dirac:1928hu} (with roots from 1928), which was placed in the theory canon, and then experimentally discovered by Anderson in 1932~\cite{Anderson:1933mb}. The positron could have been discovered in earlier experimental works, such as by Skobeltsyn~\cite{Close:2009} and the Joliot-Curies~\cite{Gilmer:2011}, had they been more versed in the latest BSM prospects and ready to recognize the positron. The discovery of the positron was a BSM confirmation because the prevailing standard reference theory, and physics community, of the time did not agree to its necessity.  Dirac was somewhat of a lone wolf crying that it needed to be there, which was strength enough to put it within the theory canon to be searched for and recognized when Anderson stumbled upon it.
Other examples in more recent times of BSM confirmations are quarks/partons~\cite{Bloom:1969kc,Breidenbach:1969kd},  parity violation~\cite{Wu:1957my}, and neutrino masses~\cite{Hatakeyama:1998ea,Fukuda:1998ah,Garisto:1998}. 

Regarding neutrino masses, it was an implicitly held view for decades that the neutrino should have zero mass and thus the SM with massless neutrinos was the default position defining the SM. As Ramond puts it, ``In fact neutrinos are absurdly light, to the point that it was widely believed that they were massless"~\cite{Ramond:2019fsr}. It was thought that since the neutrino mass should not be so much lower than the electron mass if there existed a right-handed neutrino, the explanation for the smallness of the mass then must be the consequence of the right-handed neutrino simply not existing, thereby disallowing any pairing with the left-handed neutrino to achieve a mass term. Neutrino masses would imply the need to add to the SM either a dimension-five operator, a family of right-handed neutrinos,  or more. Thus, all theories with neutrino masses were BSM theories, and began to populate the theory canon. These included theories involving a myriad of ways to naturally explain why neutrino masses are nonzero but tiny compared to other massive elementary particles in the SM~\cite{King:2003jb}. 

Confirmation that neutrinos definitively had non-zero mass occurred in 1998~\cite{Hatakeyama:1998ea,Fukuda:1998ah,Garisto:1998}, thus excluding the standard reference theory of zero masses and consequently adjusting/redefining a new SM that incorporates neutrino masses. The discovery did not happen by accident as it required tremendous investment in state-of-the-art equipment to make the BSM confirmation discovery. 

\xsection{Exclusion discoveries}

Confirmation discoveries are not possible without an experiment having the capability of exclusion. The capacity of an experiment to have exclusion discovery is a necessary, albeit not sufficient, condition for a confirmation discovery. Furthermore, with a theory canon in hand, it is possible to carry out an assessment to determine if an experiment indeed can guarantee exclusion discovery. It is arguably the duty of all resource-intensive experiments to meet the standard of guaranteed exclusion discovery. As will be discussed below, pursuing such a capacity is likely only to be positive, with no inadvertent negative side effects for a priori capacity to falsify the SM or make revolutionary discoveries, which are two complementary forms of discovery that do not require an experiment to have guaranteed exclusion abilities.

There are four main sub-categories of exclusion discovery. SM locus exclusion, which seeks to exclude a locus of BSM-inspired points within the SM parameter space. SM falsification, which seeks to find evidence that the SM is inadequate to account for all experimental data within its presumed domain. BSM exclusion, which seeks to exclude regions of parameter space within a BSM theory in the theory canon. And BSM falsification, which seeks to falsify a BSM theory, by excluding its entire parameter space. Each of these will now be discussed in turn. 

\subsection{SM locus exclusion}
\label{sec:SMlocusexclusion}

In sec.~\ref{sec:SMlocusconf} above we noted that BSM-motivated considerations can lead one to predict that future experiment would narrow the experimentally allowed region within the SM parameter space to a small locus of points. We discussed how quasi-confirmation and confirmation of a SM locus could develop in the course of experimental work. It is equally of interest to note that a SM locus could be excluded upon further experimental investigation. That is the subject of this section. 

\begin{figure}[t] 
\begin{center}
\includegraphics[width=0.47\textwidth]{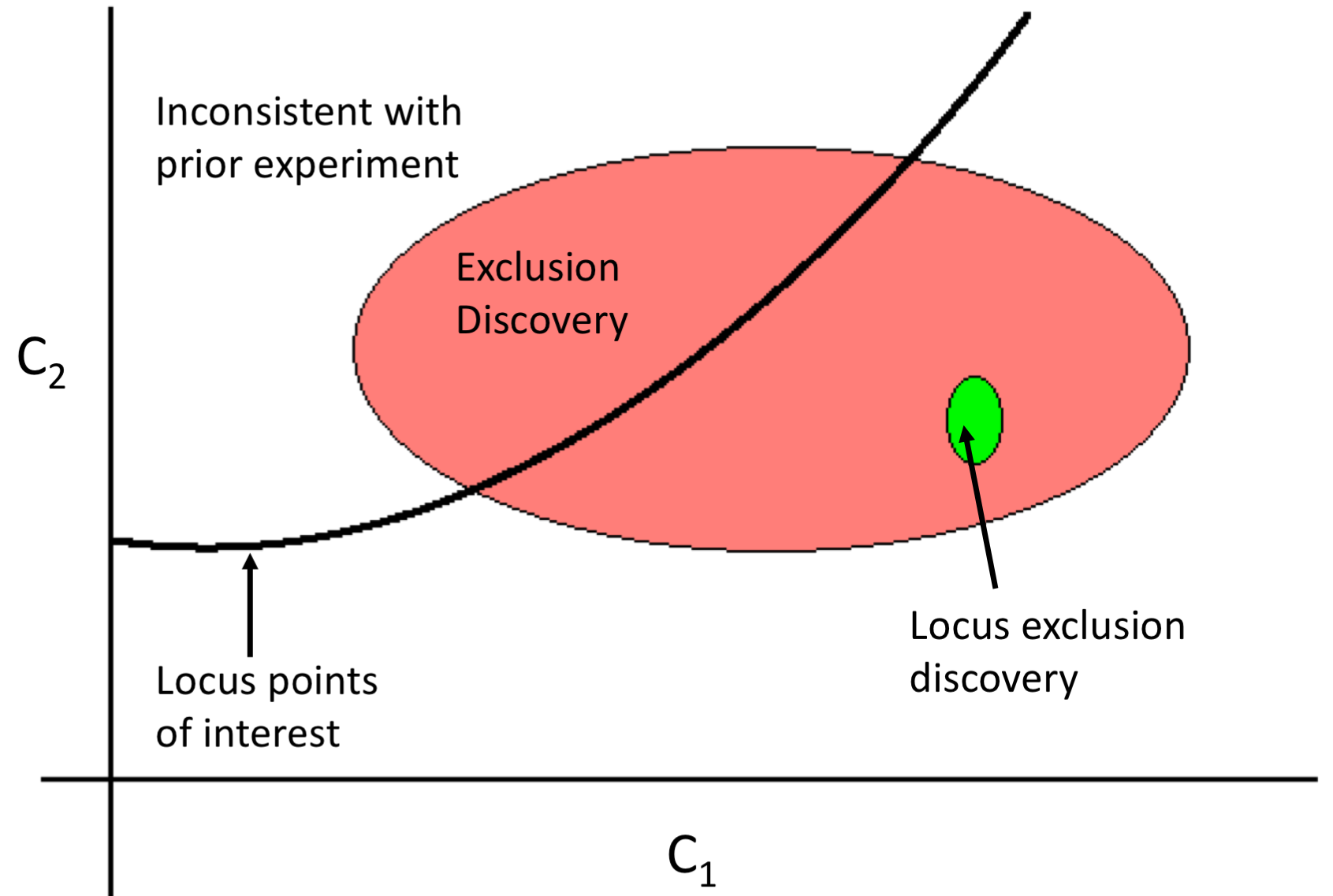} 
\caption{When the SM has parameters $(C_1,C_2)$ and a BSM theory predicts a locus points of interest (black line), a locus exclusion discovery occurs when future experiment excludes all points that lie on the locus points of interest region. This is illustrated by the experimentally allowed green region far from the locus points of interest line.}
\label{fig:Slide6}
\end{center}
\end{figure}

Recall from Fig.~\ref{fig:Slide4} the situation of an hypothesized relationship between SM parameters $c_1$ and $c_2$ (solid black line) which is initially (at $t_0$) consistent with, say, the 95\% CL region obtained by experimental measurements. After some time (at $t_1$), let us suppose that the experiment has reduced its errors significantly and is now in position to re-test whether the hypothesized relation is still viable. In this case, the 95\% CL allowed region at $t_1$, depicted in Fig.~\ref{fig:Slide6}, is the small green region, far away from the solid black line. Thus, the hypothesized relation -- the locus of points that satisfy the relation -- is excluded.  This we call the ``locus exclusion discovery." 

A locus exclusion is only a meaningful discovery if it is considered part of the theory canon, meaning that the relation was of high interest to scientists for defensible reasons. In the history of particle physics there have been many interesting locus exclusion discoveries.  One example was the hypothesized Veltman Higgs criterion~\cite{Veltman:1980mj}, which was postulated to be satisfied to control quadratic diverges of the SM Higgs sector. The criterion states that 
${\rm Str}\,{\cal M}^2(\Lambda_V)=0$ at some scale $\Lambda_V$ where the quadratic divergences are required to cancel. ${\rm Str}$ is the super-trace ($-1$ for fermions and $+1$ for bosons) over masses of elementary particles in the SM. One subtlety is to know exactly what scale $\Lambda_V$ one should evaluate this criterion. If one chooses $\Lambda_V=M_{\rm Pl}$, which is the highest known putative fundamental scale and thus where quadratic diverges would be most violently destabilizing, the condition predicts the Higgs mass to be $M_h\simeq (135\pm 2.5)\, {\rm GeV}$,  which is now excluded by more than $3\sigma$~\cite{Degrassi:2012ry}. The SM locus of Higgs masses predicted by  lower Veltman criterion scales $\Lambda_V<M_{\rm Pl}$ are excluded by even higher significance. Thus, experiment has made a locus exclusion discovery.

An example of an extremely important locus exclusion discovery was the determination that the cosmological constant is not zero. For many years it was thought that whatever solved the cosmological constant problem probably made it zero rather than a small but non-zero number, whose scale would be very hard to justify. It was a vague notion, since quantum gravity was and still is too difficult to make such predictions, but it was a qualitative possibility that was attractive. Thus, we can say that $\Lambda_{CC}=0$ was a locus of high interest in particle physics and cosmology. Excluding it would be a significant discovery by any definition of the word discovery. In 1998 that is what happened, when the experiments searching for an accelerating universe through supernova candles discovered evidence that the cosmological constant indeed cannot be zero, thus excluding that locus point~\cite{Perlmutter:1997zf,Riess:1998cb,Perlmutter:1998np}. This locus exclusion discovery remains one science's most significant discoveries of the last quarter century.

Finally, there is a third example that is interesting to discuss from the world of neutrino physics. In order to understand the masses and mixing hierarchies among the three generations of neutrinos one can introduce discrete flavor symmetries based on finite groups, which when neutrinos are assigned different representations under those groups give rise to characteristic mixing angle predictions. For example, a famous leading-order prediction for the neutrino mixing angles is so-called tri-bimaximal mixing~\cite{Harrison:2002er}:
\beq
\label{eq:bimaximal}
\sin^2\theta_{23}=\frac{1}{2},~~~~\sin^2\theta_{13}=0,~~~~{\rm and},~\sin^2\theta_{12}=\frac{1}{3}.
\eeq
This forms a locus point of prediction, which early neutrino data was consistent with. However, the present data~\cite{Bettini:2018} for neutrinos is 
\beq
\sin^2\theta_{23}=0.481-0.570,
~~~~\sin^2\theta_{13}=0.0207-0.0223,
~~~~\sin^2\theta_{12}=0.291-0.318
~~(1\sigma~{\rm ranges})\nonumber
\eeq
assuming normal ordering hierarchy of neutrino masses. Thus, present data 
is not consistent with the locus of eq.~\ref{eq:bimaximal}, and experiment has made an important locus exclusion discovery. Of course, most theories of neutrinos give rise to leading order estimates, as was stated for our prediction of eq.~\ref{eq:bimaximal}. This means that the measure-zero point of the prediction is unlikely to be exactly correct, but rather only correct to within leading order, and the ``true point" is in the neighborhood. One can define that more formally by declaring a locus of points that is within $\delta=0.1$ of the values on the right-hand side of eq.~\ref{eq:bimaximal}, or some other value of $\delta$, and then assess whether there has been a locus exclusion discovery or not. In the present case, most would agree that indeed tri-bimaximal mixing has been excluded and that a true locus exclusion discovery has been achieved by neutrino experiments.

\subsection{SM falsification}

A strong form of exclusion is the total exclusion of the SM. This is SM falsification. It is achieved by recognizing that no observable within the domain of the SM can be accommodated by any point within its parameter space. One example of how this could happen is if it were found that the $Z$ decays into $b$ quarks occur too often, in violation of precision SM predictions. In general, SM falsification  is not a question, despite simple appearances, of a single observable deviated from the SM, since one can always choose a set of parameters to make a single observable match expectations. Rather, it is a question of whether a global analysis of all measured observables within the domain of the SM ($\sigma(WW)$, $\Gamma(Z\to e^+e^-)$, $m_{\rm top}$, $m_W$, $A_{\rm FB}^{t\bar t}$, etc.)\ are compatible with at least one point in its parameter space.

The only way to accomplish SM falsification, if it is possible at all, is through extensive measurements of SM-targeted observables along with extensive theoretical work that enables the comparison of experiment with SM expectations. Thus, it puts a high premium on SM analysis development within the theory community, and the growth of experimental observables pursued and the improvement of existing experimental measurements.

It is hard to imagine anybody arguing against the importance of activities that attempt to stress-test and falsify the SM. Discovery of SM falsification would be quite momentous. It is often thought that the evidence for dark matter, coming from a variety of sources such as rotation curves of galaxies and cosmic microwave background observables, is evidence that the SM is falsified already. That is perhaps a too strong declaration since the theoretical description that provides dark matter theories may be found to be complementary add-ons to the SM rather than something that rips apart the fabric of the SM and pieces it together in a new structure. That is why the SM is still considered the standard theory within high-energy particle physics despite this strong evidence that the SM cannot be complete. Thus, researchers look for other ways to falsify the SM within the presumed domain of applicability, such as high-energy interactions of its quarks, leptons, neutrinos, gauge bosons and Higgs boson.

The worthy task of falsifying the SM can at times be confused with the notion that all thinking, both experimentally and theoretically, must be purely SM-based, with no reference to any BSM notions. To those who hold strongly to this ``signalism" viewpoint,  BSM theories are abhorrent and should be banished from scientific discussions of high-energy physics. In sec.~\ref{sec:signalism} a discussion is given on the risks of attempting to pursue discovery entirely through focus on SM falsification with no reference to BSM theories. It will be concluded there that although SM falsification is very important, and SM-based theory work and experimental work is unambiguously required for progress, SM falsification is likely to be more efficiently achieved as a byproduct of the pursuit of BSM exclusion/confirmation.

\subsection{BSM exclusion}

Consider an experiment that has the potential to make a BSM confirmation discovery. If after operating for some time a confirmation discovery does not happen, the experiment usually can place constraints on the parameter space of the BSM theory. Constraints identify exclusion regions of the theory. Although frontier experiments are hard to construct and are not commonplace, it is nevertheless common that their usual products are exclusion discoveries. Indeed, exclusion discovery generally must precede confirmation discovery.

\begin{figure}[t] 
\begin{center}
\includegraphics[width=0.47\textwidth]{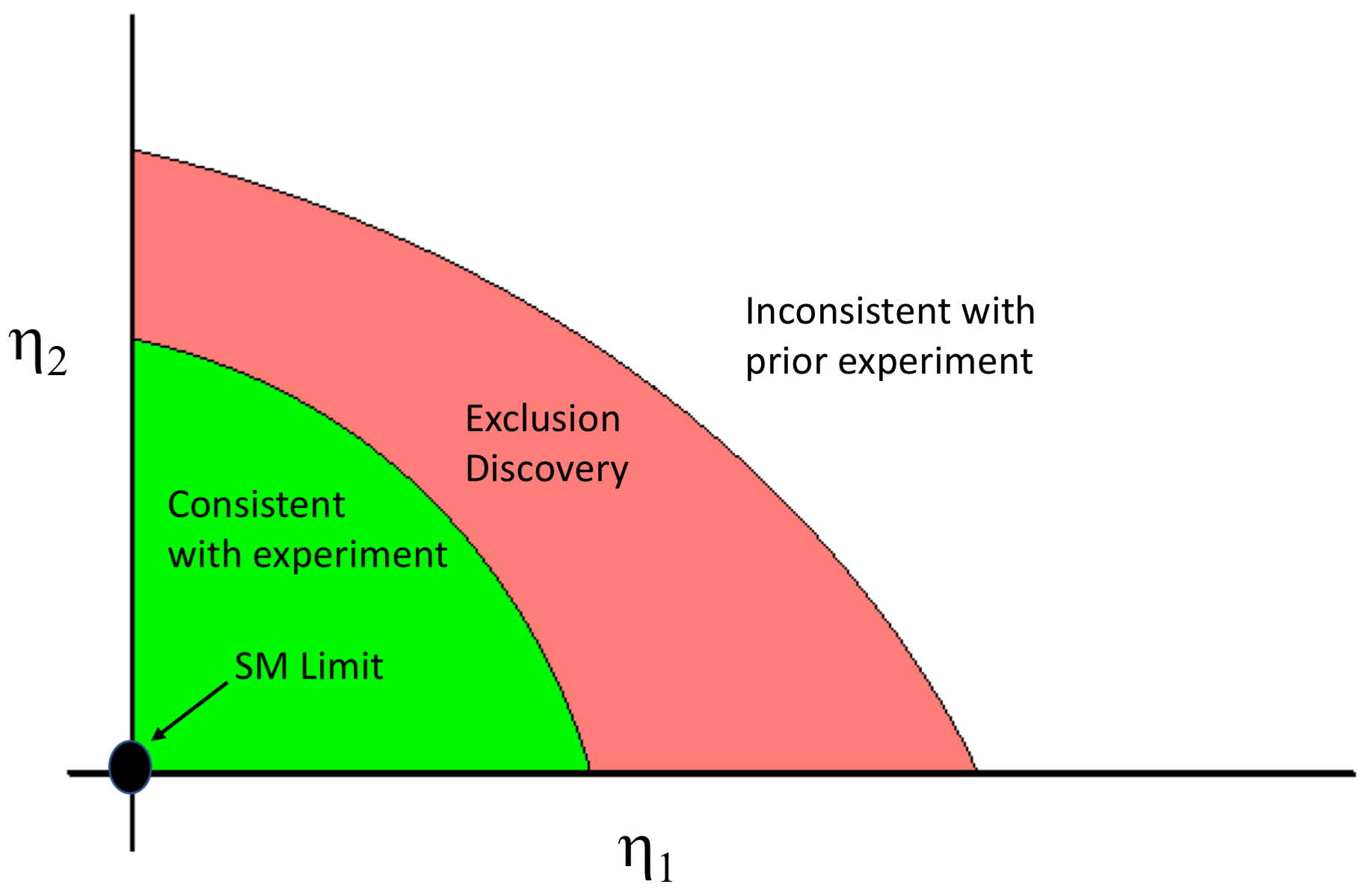} 
\caption{A BSM exclusion discovery occurs when a BSM theory has a region of its parameter space eliminated (red region) by experiment, leaving behind a smaller allowed region (green region) which is connected to the SM limit point of $\eta_{1,2}\to 0$ where all predictions of the BSM theory are indistinguishable from those of the SM.}
\label{fig:Slide2}
\end{center}
\end{figure}

Fig.~\ref{fig:Slide2} gives a representation of exclusion discovery within a decoupling BSM theory. In the figure the SM is at the origin. As experiment gathers more data the exclusion line collapses inward, and the region between the old exclusion line and the new exclusion line is the exclusion discovery. It is to be legitimately called a discovery since there was a priori potential for the BSM theory to be confirmed within that region and there is significant new information about that region in the theory canon that was hitherto unknown. Exclusion discoveries can occur simultaneously with confirmation discoveries, as Fig.~\ref{fig:Slide3} illustrates, even though the BSM confirmation result of that case would be the primary news trumpeted. Because of the simple necessitating role that exclusion must play in even confirmation discoveries, guaranteed capacity for exclusion discovery should be the standard by which future proposed experiments are judged. If they wish to be transformative experiments and make discoveries, they will have to extend exclusions beyond what the sum total of all prior experiments could do. 

As an example, the LHC has made numerous exclusion discoveries in addition to its celebrate SM confirmation discovery of the Higgs boson. It has excluded large regions of parameter space (i.e., of the theory canon) for minimal supersymmetry and minimal models of composite Higgs. See Figs.\ 12 and 13 of ref.~\cite{Aaboud:2018ujj} and Figs.\ 2-5 of~\cite{Arbey:2015exa} for schematic representations of the exclusion discoveries within these theories. It has also made exclusion discoveries for theories with leptoquarks, extra gauge symmetries, extra spatial dimensions, etc. 

The frequent recitations of exclusion discoveries are at times cast as pronouncements of failures~\cite{Browne:1993},  but of course they are necessary and must occur frequently for progress to be assured. Instead, they should more rightly be viewed as successful executions of experimental work that had capability for such exclusion discoveries. 
Excluding enormous swaths of the theory canon is a major achievement of experiment, as one would quickly realize if they tried to do that at home.
A vibrant and rich BSM theory canon means there will be many individual disappointments, but effort is sustained and rewarded when viewed as a collective search party exploring uncharted territory for deeper hidden knowledge.

In short, exclusion discoveries are important discoveries that are necessary and expected when experimental science progresses. The existence of an exclusion discovery signifies that a BSM confirmation discovery was possible,  highlighting the experiment's value. Furthermore, well-articulated exclusion discoveries raise the bar for future experiments to have discovery potential, allowing us to question sharply experimental plans and projects that cannot demonstrate future guaranteed discovery potential. 

\subsection{BSM falsification}

There are times when experiment makes sufficiently strong exclusions that an entire BSM theory (i.e., a BSM theory's entire parameter space) is falsified out of the theory canon due to its inability to be empirically adequate. An example of this is the minimal fourth generation model, which postulates that there is another generation of fermions in addition to the three generations that we already know, and that these fermions all have degenerate (or nearly degenerate) mass. Careful precision measurements at LEP  were enough to rule out this theory~\cite{Amsler:2008}. This can be called a BSM falsification discovery, and is an extreme version of exclusion discovery.

\begin{figure}[t] 
\begin{center}
\includegraphics[width=0.47\textwidth]{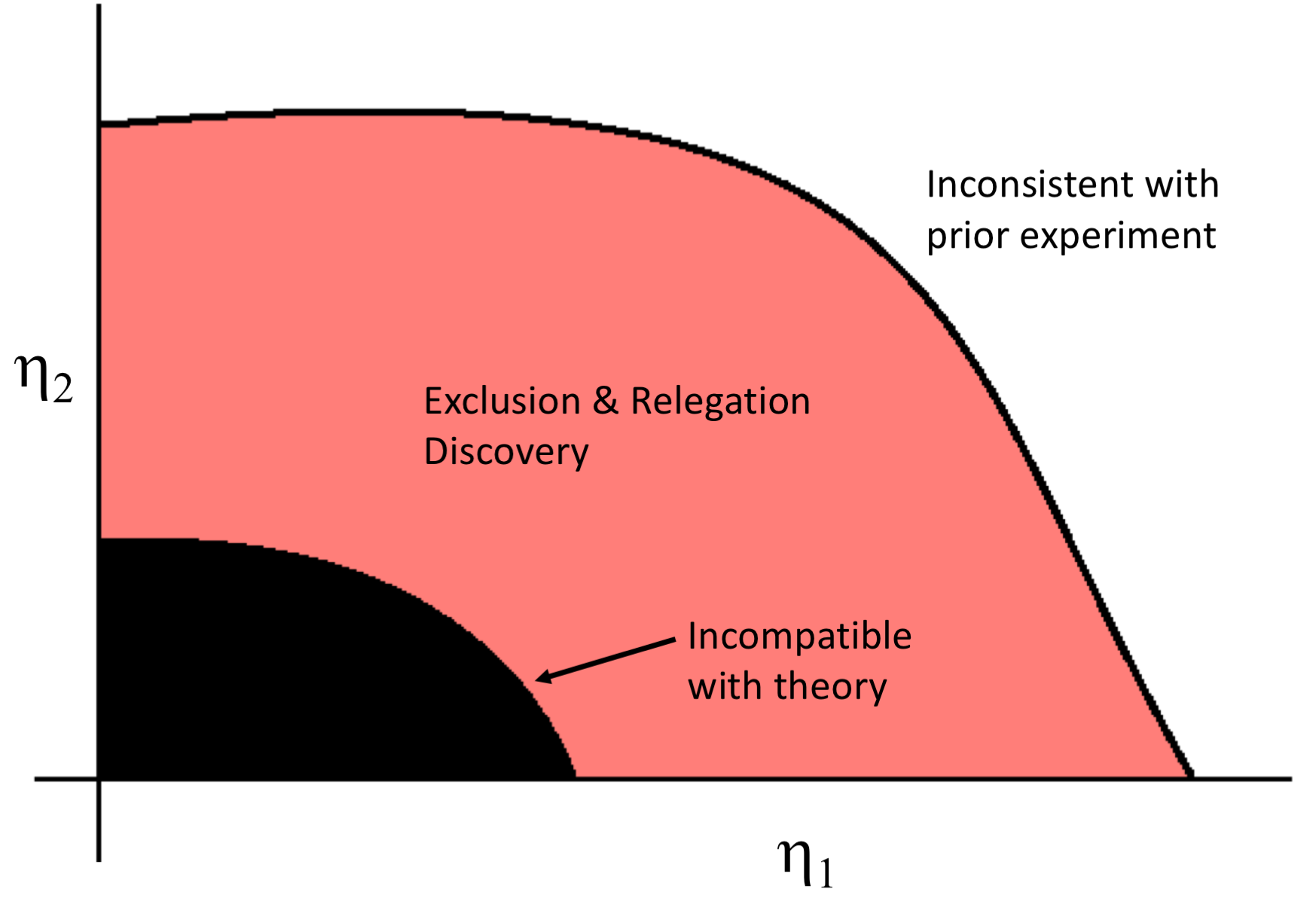} 
\caption{A BSM falsification discovery occurs when a BSM theory has its entire parameter space eliminated (red region) by experiment, leaving no allowed region behind. If all observables are still consistent with the SM, a BSM falsification can only occur if the BSM theory has no SM limiting point, either by construction of the theory, or by eliminating a region around the SM limiting point of $\eta_{1,2}\to 0$ by non-empirical methods, such as by imposing limits on maximum tolerated finetuning of parameters.}
\label{fig:Slide9}
\end{center}
\end{figure}

Another BSM falsification discovery made by Tevatron alluded to in the introduction is that of minimal no-scale supersymmetry with neutralino dark matter. In this theory, there is an upper bound on the value of $m_{1/2}$, above which the neutralino cannot be the dark matter. In a plot with decoupling parameter(s) where the SM is at the origin, this would be equivalent to eliminating by theory construction the parameter space in the neighborhood of the SM.  Thus, the theory does not have a decoupling SM limit point within its definition that would give it experimental safety against continuing exclusion discoveries. At some point the exclusions run out of real estate and the theory cannot be accommodated. At this point a BSM falsification discovery has been made. A schematic representation of a BSM falsification discovery is given in Fig.~\ref{fig:Slide9}, where a transformative theory has turned a once green region (i.e., consistent with experiment) entirely into a red region, such that there is no longer any experimentally allowed region for the BSM theory. 

BSM falsification discoveries can be controversial especially if a theory's parameter space is reduced by non-empirical methods. For example, some might say that the LHC experiments have made a BSM falsification discovery by completely excluding supersymmetry. 
Another example is the purported falsification of the minimal grand unified theories, both supersymmetric~\cite{Murayama:2001ur} and non-supersymmetric~\cite{Senjanovic:2009kr}. However, these BSM falsification discovery  require a strong cut on parameter space by non-empirical means, such as not taking into account the possibility of higher dimensional operators or taking naturalness and finetuning criteria very seriously and assuming a rather aggressive (i.e., not conservative) low cutoff value of finetuning to be acceptable~\cite{Wells:2018sus,Wells:2018yyb}. 

Nevertheless, if the community, or some respectable fraction of the community, deems these extra-empirical conditions to be required for (non-)admission the theory canon, then so be it. However, the BSM falsification discovery should not be declared without a very precise description of exactly what theory is being expelled from the theory canon, including whatever additional non-empirical criteria applied. This is often lacking.

One should emphasize that although naturalness and finetuning considerations were invoked originally because it was thought that the next stage theory is likely to be natural, it has also served, perhaps subconsciously, as the only means by which a BSM falsification discovery could be made for many theories. Without it there can be no exclusion  of supersymmetry, or most of its invariants, for example, according to our current understanding of the theory and the data. This is true of all decoupling theories -- composite Higgs, extra dimensions, supersymmetry. Without some way to expel a finite region around the SM decoupling limit point(s), there is no way to ever falsify them when data stays consistent with SM. One just makes exclusion discoveries (or confirmation discovery) closer and closer to the SM limit point(s) in the parameter space. Naturalness also has served indirectly  a very useful practical role in that it has encouraged physicists to think much more about experimentally accessible phenomena, since the decoupling region without experimental consequences is an anathema to a natural BSM theory\footnote{For this reason, and others, it is baffling why anybody who cares deeply about theorists focusing on theories that are accessible to experiments should think it destructive to science progress that a researcher is encouraged by the naturalness criteria when theory model building. On the other hand, if theorists had become enamored with the ``principle of anti-naturalness," where every new theory had to be highly finetuned for some reason, and thus typically out of reach of every conceivable experiment, that would be a significant concern to science progress. Thankfully, that never happened.}.

Some of the most interesting experiments are those that have made BSM falsification discoveries. ``Ruling out" theories has large impact in how science progresses. Ruling out theories with the simple QCD axion, additional light neutrinos, degenerate fourth families, minimal versions of supersymmetric theories, minimal technicolor theories of electroweak symmetry breaking, minimal top-quark condensate theories, minimal $SU(5)$ grand unified theories, etc., have all set the field in different directions which were by definition more productive. Each of these BSM falsifications took prodigious experimental skill and resources to achieve, and each has had tremendous impact on high-energy physics. To increase the meaning and impact of experiments, articulating precisely all the BSM falsifications that it has achieved within the theory canon is a useful endeavor. This may involve creative categorical parsings of parameter spaces within bigger frameworks (supersymmetric, compositeness, etc.) but the results expressed are meaningful and powerful, and set the standards above which future experiments must achieve.

\xsection{Revolutionary discoveries}

There are times when experiment has results that are in conflict with every theory within the theory canon. Such discoveries can be called ``revolutionary discoveries" since it annihilates the entire theory canon and requires one to start anew in model building that takes into account the new results. It goes without saying that the experimental result must be solid and reproducible and beyond reproach in order to claim a revolutionary discovery.

Perhaps ``annihilating" the theory canon is too strong, since the old standard theory is likely to still be of use in a domain restricted compared to what is once was before the revolutionary discovery. For example, after the muon discovery it is understand that the total cross-section of $e^+e^-$ annihilations is higher than the old QED results for center of mass energy greater than $2m_\mu$, since $e^+e^-\to \mu^+\mu^-$ contributes now in addition to the standard $e^+e^-\to e^+e^-,\gamma\gamma$, etc.\ results of QED. However, the total rate for $e^+e^-\to e^+e^-$ is not significantly changed\footnote{Nevertheless, there is a change, albeit tiny, since very precise measurements would be sensitive to quantum loops of virtual muons in the photon propagator mediating $e^+e^-\to e^+e^-$.} due to the discovery of the muon and thus the old theory is still useful. Indeed, the muon can be merely added to the QED lagrangian in a way directly analogous to the electron except that its mass is higher. 

Nevertheless, what is clear is that no theory as it stood before in the theory canon survives a revolutionary discovery, by definition.  In the case of the muon discovery, that new theory was not terribly difficult to devise since the muon interactions were so similar to the electron interactions. However, it was a new theory, and it is reasonable to conclude that the experimental discovery was revolutionary according to standard connotations of the word.

Some revolutionary discoveries lead to a new theory canon that is initially all incorrect. An example of this is the anomalous perihelion procession of Mercury. Once the experimental result was established by Le Verrier, it was thought that the theory needed to change or the objects that were part of the theory description (sun, planets, asteroids, etc.) needed to change. The most compelling idea was a new planet  between Mercury and the Sun, but later experiment did not find it. New ideas that invoked more finely grained objects in dust belts that could not so easily be found by experiment were then invoked, but experiment ruled out parameter space more and more finely for such ideas~\cite{Roseveare:1982}. What ultimately worked was an entirely new theory within the theory canon -- General Relativity -- which explained that result. It also made an additional non-trivial prediction that no other theory in the canon made, that of  the bending of light. When a confirmation discovery was made for the bending of light, General Relativity sat most prominently in the theory canon for describing gravitational phenomena. In this history, the discovery of the anomalous perihelion precession of Mercury was a ``revolutionary discovery" and the discovery of the bending of light was a ``confirmation discovery." For those who were not convinced of general relativity yet, it was a ``BSM confirmation discovery" while for others who were already convinced it was a ``SM confirmation discovery." In any case, it was a discovery for all.

There are many other examples of revolutionary discoveries in the recent history of physics. Events with missing energy in nuclear $\beta$ decays were unexpected and thus were revolutionary discoveries. The missing energy was ultimately explained by the existence of neutrinos, which was confirmed~\cite{Pais:1986}. To some it was a BSM confirmation discovery since the discovery of neutrinos was preceded by the theoretical prediction of its existence, whereas others would view it as a revolutionary discovery since the original experimental result annihilated the entire theory canon that had existed at the time, although it took physicists some time to realize that. The discovery of the acceleration of expansion of the universe would qualify to some as a revolutionary discovery since in so many people's minds it was inconceivable that the vacuum should have a positive cosmological constant, which is what the results imply. To others it was a ``locus exclusion discovery," excluding zero cosmological constant value, as was discussed above in sec.~\ref{sec:SMlocusexclusion}.

In summary, a revolutionary discovery happens when the entire theory canon is falsified by unforeseen phenomena.  Revolutionary discoveries cannot be guaranteed nor even anticipated. They happen ``out of the blue." Nevertheless, revolutionary discoveries do happen, and an important remaining question of this analysis of discovery is whether focus on BSM exclusion, which has been argued to be the only path to assured  discovery, dims the prospect for revolutionary discovery, which is never assured and which by definition takes place outside of the entire theory canon and its array of BSM theories. That is the topic of the next section.

\xsection{Signalism: risks of pursuing discovery without BSM context}
\label{sec:signalism}

The discussion above points to the utility of pursuing discovery with a BSM-oriented approach. Historically there is much agreement on this approach, although the language by which it is phrased may be different. For example, as A.P. Aleksandrov reports, ``[Euler] himself believed that science progresses via conjectures, by successively rejecting less accurate conjectures in favor of more complete ones"~\cite{Bogolyubov:2007}. And Feynman said,
\begin{quote}
In general we look for a new law by the following process. First we guess it. Then we compute the consequences of the guess to see what would be implied if this law that we guessed is right. Then we compare the result of the computation to nature, with experiment or experience, compare it directly with observation, to see if it works. If it disagrees with experiment it is wrong. In that simple statement is the key to science.~\cite{Feynman:1965}
\end{quote}
These two quotes clearly highlight the importance of the BSM confirmation/exclusion approach to discovery, in our language. 

Nevertheless, there creeps in a counter-sentiment to the BSM confirmation/exclusion approach to discovery, which Feynman also brings up in the same lectures wherein the above quote was delivered:
\begin{quote}
This [the BSM confirmation/exclusion approach, in our language] will give you a somewhat wrong impression of science. It suggests that we keep on guessing possibilities and comparing them with experiment, and this is to put experiment into a rather weak position. In fact experimenters have a certain individual character. They like to do experiments even if nobody has guessed yet, and they very often do their experiments in a region in which people know the theorist has not made any guesses. For instance, we may know a great many laws, but do not know whether they really work at high energy, because it is just a good guess that they work at high energy. Experimenters have tried experiments at higher energy, and in fact every once in a while experiment produces trouble; that is, it produces a discovery that one of the things we thought right is wrong. In this way experiment can produce unexpected results [revolutionary discoveries in our language], and that starts us guessing again.~\cite{Feynman:1965}
\end{quote}
From these two quotes of Feynman above, within the same lecture we see two different approaches to pursuit of discovery: the BSM confirmation/exclusion approach  and the no previous ``guesses"  approach to experiment\footnote{One is tempted to call this latter approach the ``shut up and build" approach to experimental science.}. Feynman does not resolve that tension in his lecture, or comment further on the relative merits of each. Perhaps a healthy balance of both approaches is the best path, one might wish to think. It will be argued here that a primary orientation to BSM exclusion/confirmation is superior, while at the same time it does not diminish the prospects for surprises. Seeking pure surprises outside the context of a BSM orientation is a riskier endeavor.

To develop and define the tension further, let us first acknowledge that, as with Feynman's ``experimenters" in the previous quote, there has been and is now a current within particle physics that is reluctant to embrace theoretical speculations except in times of crisis when no theory is empirically adequate at all. For example, when $\beta$-decay was shown to have what looked to be violations of energy conservation, speculation as to what could cause that signal was acceptable. When new signatures were found in cosmic rays, which ultimately led to our conceptualization and discovery of the muon, speculation as to what the origin was of that signature was acceptable. In contrast, speculations that are not in the service of an extreme experimental crisis are considered by some to be ``philosophy" that has very low efficiency in revealing truth. To such individuals, it is much better to become divorced entirely from BSM consideration and approach experimental studies in a ``model independent" way or in a ``signal-based way." The desire, in other words, is to seek SM falsification or revolutionary discoveries with no reference or consideration of BSM rationales. A key assumption of this mindset is that focus on BSM exclusion/confirmation discoveries derails experiment from the more productive focus on SM falsification or revolutionary discovery pursuits, which then as a byproduct creates a crisis --- the standard theory being unambiguously incompatible with experiment --- and thus opens the door to focused, fruitful, BSM theory. In the absence of such a crisis there should be no BSM theory work; there should be only SM theory work so as to more effectively identify when the SM has been falsified and when a true revolutionary discovery has indeed been made.
This approach can be called ``signalism" with its emphasis on initiating and studying signals, then comparing them with SM expectations, all at the exclusion of any BSM theory context.

Again, it will be argued in this section that signalism poses a significant risk to discovery of every kind, including its targets of SM falsification and revolutionary discoveries. One of the themes that will be explored below is that no matter how hard one tries, escaping BSM considerations is impossible if one wants to give meaning and context to the SM and have any rational basis for future experiment. As we will discuss, to make an argument for a future experiment or experimental analysis is necessarily to engage in BSM theory assessment, whether one recognizes it or not\footnote{Analogs to this ``you cannot escape speculative theory" argument can be found everywhere in intellectual pursuits, as far and wide even as literary theory: ``Hostility to theory usually means an opposition to other people's theories and an oblivion of one's own"~\cite{Eagleton:1996}.}.

Let us start by giving some of the arguments for and against the proposition that BSM-exclusion/confirmation approach to planning and executing experiment reduces prospects for revolutionary experimental discoveries. 
On one side of the argument, one could claim that focus on BSM exclusions within the theory canon promotes a more narrow set of experiments aimed at the narrow set of phenomena that the BSM canon theories predict, leaving out searches for the vastly greater array of new phenomena that a less theory-laden approach to experiment could probe. On the other side of the argument, one could claim that the vast majority of discoveries of the last century are BSM confirmations and not revolutionary discoveries, and that with finite resources (i.e., we cannot cover/measure all conceivable phenomena one could think of anyway) the search for BSM exclusion/confirmation is the best investment. In any event, focus on BSM exclusion/confirmation is an experimental activity that {\it a priori} has arguably at least just as much chance of finding something unexpected (i.e., revolutionary) as any other conceivable approach.  

Furthermore, even if one did not like BSM-exclusion/confirmation centered approach to experimental searches and wanted to instead focus on initiating a program to maximize revolutionary discoveries, a curious contradiction develops. The moment one seeks a rational description of an investment with aims toward revolutionary discovery is the moment one unwittingly entertains nebulous ill-conceived BSM notions well outside what experts in BSM theory would think is interesting or that could possibly solve any recognized problem that the SM does not address. And as soon as an argument ensues that the established BSM experts are wrong in their assessments, and that one's vision of what is possible in BSM physics is defended, the researcher then becomes a BSM theorist, losing their revolutionary-only claim. Thus,  attempts to seek only revolutionary discoveries in research, while rebuffing any and all BSM notions, necessarily dissolves into the mystical and visionary and away from the rational. In such a universe of thought that consciously runs away from rationales  it is hard to decide whether throwing a vase off the Eiffel Tower to see what new thing might happen is better than analyzing high-luminosity LHC data. 

To further explore these issues, one approach to try to be a revolutionary experimentalist is to decide, consciously or subconsciously, that any BSM phenomena or signal (not necessarily derived from a theory) ever expressed that is incompatible with the SM but not in contradiction with prior experiment is currently viable, and experiment could be chosen to pursue a randomly chosen (because of limited resources) new signals identified among them, thereby eliminating ``theory bias." However, this approach is yet another BSM theory, which requires the researcher to believe that a randomly chosen new phenomenon/signal identified by humans that is not possible within the SM, but which is not yet excluded by experiment, is more likely to be manifested by nature than any phenomena derived from theories constructed for the purpose of solving outstanding problems in the SM. There is no known logical justification for such a signalist belief.

A closely related  line of argument would be  to claim that pure focus on BSM exclusion/confirmation necessarily dims the prospects of revolutionary discoveries since it prohibits signalism-like approaches to science from being funded and pursued. However, the missing element of that argument is that signalism has yet to be justified, and once it is justified sufficiently to garner resources it becomes an established BSM theory, and we are back to focus on BSM exclusion/confirmation. This is not to mention that  it is hard to imagine any compelling justification for signalism, or related ideas, since the number of very odd signals possible versus those realized in nature is arguably infinite and thus the probability would be vanishingly small for selecting a good one to pursue in the search for a revolutionary discovery. This leaves standard (non-signalismic) BSM exclusion/confirmation as a preferred focus for discovery.

An example illustrating this is in the career of Nobel Prize winner Martin Perl, whose attitudes and public pronouncements can best be described as signalismically oriented, but whose activities mixed BSM exclusion/confirmation with signalism-based approaches, with its inevasible slide toward BSM justifications (see, e.g.,~\cite{Lee:2004jg}).
Perl frequently made the  statement that experimentalists should be cautious about, and perhaps even ignore, popular physics theories (i.e., the BSM theory canon), as evidenced by his 1986 essay~\cite{Perl:1986}. In that essay, he made several claims that are at odds with the thesis of this essay. For example, Perl wrote, 
\begin{quote}
Experiments based on speculative theories and with narrow goals teach us little if the answer is
no --- only that the theory is wrong or, more likely, that the parameters in the
theory need adjustment.~\cite{Perl:1986}
\end{quote}
However, there are two problems with this statement. First, he conflates ``experiments based on speculative theories" and ``experiments with narrow goals."  This is a false dichotomy on experiment. Experiments can be extraordinarily ambitious and broad and yet target ``speculative theories" (BSM theories). 

Furthermore, Perl makes the claim, in our language, that BSM exclusion discoveries ``teach us little." As opposed to Perl, we have argued here in this essay that  if the SM is not falsified and a BSM confirmation/exclusion discovery has not occurred, then we have learned very little. The primary knowledge that most experiments in history have taught us is through BSM exclusion/confirmation discoveries. It teaches us in all the ways discussed above, including creating a new threshold above which future experiments must achieve to be deemed worthy successors. In Perl's essay there is not a developed argument for what it means for experiment to teach us something, but one might infer that, in the absence of SM falsification and BSM confirmation, what it can teach us primarily in Perl's view is the recital of unadulterated experimental data, to be written on velum, metaphorically, and stored for humanity's gaze in perpetuity. However, there is nothing incompatible with this majestic view of experimental data having worth in and of itself, and the additional view that it can be utilized to monitor and characterize the status of BSM canon theories.  After all, it should never be lost that the SM is not the only theory compatible  with all data. Ironically, only a deeply held non-empirical mindset, which necessitates strong speculation, would demand intense loyalty to the SM only, and show disdain for competing empirically adequate theories in the canon.


Another closely connected question is whether focus on BSM-exclusion/confirmation dims prospects of falsifying the SM. This is especially problematic for those who adhere to the signalism viewpoint.  According to the signalism mindset, the primary path of science is to decide on the Standard Theory,  which in the case of high-energy physics is the SM, and once it is decided, one should focus entirely on falsifying it. In other words, the only activities of any value are those that stress-test the SM to the extreme with the hopes of breaking the SM, either through unambiguous statistical anomalies of SM observables or through revolutionary discoveries, such as new non-SM resonances. Only after the SM is broken should we pick up the pieces and do fancy theorizing that constructs a new SM. Any activities that build and analyze more ambitious BSM theories are terrible wastes of time and very inefficient, since more or less all ideas except at most one must necessarily ultimately crumble to dust with more knowledge, which by the way is not guaranteed anyway. For this reason, again, the only theory work should be that which computes a myriad of experimental observables to higher and higher accuracy with an aim to reducing the  allowed SM parameter space to the smallest possible volume. 

It must be acknowledged that the pursuit of SM falsification is unquestionable a coveted achievement in high-energy physics since  falsification would imply that qualitative new understanding of nature is needed to establish a new SM that accommodates the data. However,  narrow focus only on SM exclusion without the benefit of a BSM perspective risks derailing the very falsification goals it is trying to achieve. For example, a narrow SM-only focus does not imagine any ways that observables could go awry and only attempts to get the very best measurements possible to squeeze the allowed parameter volume of the SM compatible with experiment to smaller and smaller values. 

For example, a SM-only perspective could suggest that instead of venturing into an energy frontier one could only increase the intensity frontier, gathering more and more $Z$ boson decays, and more and more $W$'s and top quarks to compare with precision measurements. Measuring high-energy $e^+e^-$ or $pp$ collisions at significantly higher energy but with somewhat limited luminosity appears worthless to a SM-only perspective. Without a BSM perspective it is hard to ever imagine a strong reasoned case for going to the high energy frontier. No parameters of the SM will be measured better by such new endeavors in many cases. Yet, we know that from the perspective of the SM being an effective theory of a more extensive higher energy theory that the effects of new physics, which if seen would break the SM, often become more and more pronounced at higher energies. The development of SM-incompatible signals (SM falsification) at higher energies is due to momentum-dependence of higher dimensional operators in the effective theory, or the opening up of new particle thresholds. This is decidedly a BSM-perspective, developed by vast experience with BSM theories, and could never be divined from an unadulterated SM mindset. 

In contrast to a BSM-informed mindset, the pure SM mindset could keep one anchored forever to pursuing high statistics right around the weak scale in order to measure SM parameters better and better, with hope that one day precision statistical analysis would develop a deviation that grows over time, not even considering the possibility that a new particle or new interaction could be not far away in the energy frontier or in the frontier of new experimental methods. Again, as soon as one even contemplates a motivation to spend more money and go to higher energy because, for example, ``a new particle might be there," they have entered the BSM world whether they like it or not, and thus must face the fact that their speculative motivations to spend more money and build a costly energy frontier facility could be challenged or criticized, and thus must be justified in some way at least as plausible. In other words, they need to articulate why their BSM speculations are worth pursuing. A vibrant community of BSM scholars can then be appealed to for that task, and thus helpful in the informed co-pursuit of SM falsification, even when trying to falsify the SM is the experimentalist's only goal.

Arguments of discovery centered on BSM exclusion/confirmation vs.\ signalism abound in less august forums, such as social media posts, blogs, and letters to the editor pages of Physics Today. A particularly prominent one from the early 2000's was Harry Lipkin's {\it crie du coeur}:
\begin{quote}
I have no patience with social scientists, historians, and philosophers who insist that the `scientific method' is doing experiments to check somebody's theory. The best physics I have known was done by experimenters who ignored theorists completely and used their own intuitions to explore new domains where no one had looked before. No theorists had told them where and how to look~\cite{Lipkin:2001}.
\end{quote}
and Lincoln Wolfenstein's equally forceful retort that lists theory confirmation after theory confirmation discovery and ends with 
\begin{quote}
We do not have a theory of everything, although some of my colleagues dream of one. When new domains of energy are explored, we will not be surprised to discover that there are things in the heavens and on Earth that are not described by our present theory. Our goal, then, must be to find a more encompassing theory and design experiments to fully test it. That, I believe, is the scientific method~\cite{Wolfenstein:2001}.
\end{quote}
In the end Lipkin's signalism-oriented viewpoint, springing from a rather cursory historical analysis, emerges naive compared to Wolfenstein's BSM-oriented view. That the most consequential physics discoveries have ``ignored theorists completely," as Lipkin claimed, is rather easily countered with the simple realization, for example, that no experiment ever measured the process $e^+e^-\to {\rm SU(2)}_L$.

In summary, signalism is the desire to make SM falsification discovery or revolutionary discovery through construction of signal-based analyses that compare experiment with SM theory, while disallowing all reference to speculative BSM theories. The problem, as discussed above, is that tacking to revolutionary discoveries without BSM reference is an inscrutable exercise that can never have a rational justification\footnote{The risks of pursuing revolutionary discoveries through new experiments without any theory context allowed, which then does not allow comparisons of value with respect to prior experiments and observations, has been illustrated well recently by Caldwell and Dvali in the specific case of anti-matter gravity experiments~\cite{Caldwell:2019icl}.}.  At the same time, tacking to SM falsification discovery without BSM reference can lead to experiment pirouetting longer and longer on the same or similar experimental analyses, accruing more and more data over time as they improve SM parameter determinations with hopes of a statistical incompatibility developing, meanwhile feeling very little pressure to think of and pursue different types of experiments that test more fundamental theory lying in wait ``just beyond." That is not to say that achieving higher precision on SM observables is not worthwhile physics. However, when deciding which parameter to pursue to much higher precision, BSM insight helps decide. BSM also helps decide when results of a particular experiment and signal are good enough, and something else should be done. Focus on BSM physics is key for pruning away inscrutable or unproductive pursuits, for guaranteeing discovery (BSM exclusion/confirmation), and for  enhancing prospects of SM falsification and revolutionary discoveries.

\xsection{Gravity waves \& Higgs boson discoveries through the BSM lens}

Two of the most momentous discoveries in particle physics and cosmology in the last decade have been the Higgs boson and gravity waves. In this section these two discoveries are compared and it will be argued that the excitement for the discoveries comes not because of the rush that comes from finally seeing and confirming a standard theory feature that we have been talking about as a science community for such a very long time. Although that is of high interest, that is not the core reason why the community has demonstrated so much enthusiasm over these discoveries. Rather, the intense excitement  is to be understood as the recognition that a new era has been ushered in of present and future BSM exclusion/confirmation discoveries that were heretofore inconceivable.

Regarding gravity wave physics, let us make a few remarks about our venturing away here from traditional particle physics to discuss it.  Up until now we have mostly focused on experiments very closely tied to the SM and its BSM extension, which has lead us primarily to discuss experiments that have reproducible human-induced conditions and phenomena which then is measured by experiment. However, the discovery conceptualization discussed above and the central role that BSM is argued to have are applicable to any forefront, basic science field. This is especially applicable to cosmology which shares its intellectual domain with high-energy physics for the simple fact that early time (cosmology) means high energy (high-energy physics). Gravity wave physics has breadth across many early epochs of physics, including early universe cosmology. There are standard reference models for star formation, and standard reference models for binary merger theory, and standard reference models for cosmological evolution (radiation domination then matter domination), etc. All of these can be framed as theory canons with experiments that make exclusion/confirmation discoveries with respect to them.

The discovery of gravity waves came with much fanfare, even though they were completely expected. The last reasonable scientist to be unsure if gravity waves really existed was Einstein in 1936~\cite{Kennefick:2005}. Since then there has been no serious questioning of their existence within the physics community. Thus, one might be tempted to say that the discovery of gravity waves by LIGO experiment was merely a boring SM feature confirmation discovery and did not advance science more than what we already knew with very high confidence. 

So what exactly is it that is so exciting about the LIGO detection of gravity waves if it was a merely confirmation of what we already were sure of? The answer is that it opens the door to a vast array of guaranteed BSM exclusion/confirmation discoveries in the future. It is the reason why the ``SM discovery" of gravity waves did not end the field, but rather marked the start of a new discovery era of BSM exclusion/confirmation discoveries. We now have a better idea of what the rate of the gravity waves are and we have reached the experimental threshold where we can actually measure them and plot their waveforms. Again, it is not the discovery of gravity waves themselves that is so exciting, but its implication for future discovery. Already, there are innumerable studies of BSM ideas for early universe cosmology and astrophysics that will be tested and excluded/confirmed, leading to a vastly deeper understanding of nature. These ideas of testing new physics~\cite{Caldwell:2018giq} are wide-ranging, including probes of first-order phase transitions~\cite{Baldes:2018nel}, tests of early universe equation of state~\cite{Cui:2017ufi,Cui:2018rwi,Redmond:2018xty}, probes of axions physics~\cite{Poulin:2018dzj},  tests of and early kination phase in the early universe, and much more, all made possible by the discovery of gravity waves.

In contrast, the Higgs boson discovery came with great fanfare for two reasons. First, unlike gravitational waves, the Higgs boson was considered by many to be a speculative possibility even moments before its discovery was announced in 2012~\cite{Wells:2018nwj}. Thus, the discovery in and of itself was much more significant to science than the discovery of gravity waves in and of themselves. 

Equally important, and more directly analogous to the gravity wave discovery, the discovery of the Higgs boson heralded the dawn of a new era of BSM exclusion/confirmation discovery. Careful measure of the Higgs boson mass enables tests of the possible composite nature of the Higgs boson~\cite{Carena:2014ria}.  It can test alternative forms of the electroweak symmetry breaking potential, including $|H|^6$ terms that could enable a first order phase transition~\cite{Grojean:2004xa}. It opens a portal to hidden worlds made possible by the only dimension-two operator in the SM that is gauge invariant and Lorentz invariant ($|H|^2$)~\cite{Schabinger:2005ei,Patt:2006fw}. It can test supersymmetric theories since Higgs mass is computable from the supersymmetric spectrum~\cite{Draper:2016pys}. It enabled tests of cosmological ideas, including stability of the universe~\cite{Degrassi:2012ry} and even inflationary theories~\cite{Bezrukov:2007ep}. Precision Higgs boson production and decay measurements  will enable innumerable exclusion and confirmation discoveries of BSM ideas through SM locus confirmations (e.g., predicted ratios of Higgs branching ratios) and BSM exclusion/confirmations (e.g., exotic decays of Higgs boson), all of which were made possible for the first time by reaching the experimental sophistication of copiously producing and studying the Higgs boson~\cite{Curtin:2013fra,Dawson:2018dcd}. This richness of BSM exclusion/confirmation discoveries made possible by the discovery of the Higgs boson forms the center of the physics cases for many new facilities, most particularly the ILC~\cite{Fujii:2017vwa} and CLIC~\cite{Abramowicz:2016zbo} which have stages that are designed exclusively for these opportunities.

There are other SM feature discoveries that are being made continuously that have not garnered as much attention. For example, recently LHCb has discovered two new resonances, 
$\Sigma_b(6097)^+$ ($buu$ boundstate) and $\Sigma_b(6097)^-$ ($bdd$ boundstate)~\cite{Aaij:2018tnn}. There are any number of reasons why such a discovery is of high value within particle physics. Above all, it is learning more about nature -- the discovery and confirmation of the kinds of structures that are allowed. It enables additional data to aid development of computational techniques. It might one day be useful in testing new physics ideas in ways that are not currently anticipated. Nevertheless, it does not have the same panache as hadronic resonance discoveries of years ago\footnote{As Martin Perl put it, ``20 years ago the discovery of an additional hadronic resonance was an important event in our world; now such a discovery gains no recognition beyond a new entry in the particle data tables"~\cite{Perl:1986}.}.

Why are discoveries of new hadronic resonances not met with tremendous fanfare in the science community? The reason is that they no longer are thought to have much impact on BSM exclusion/confirmation discoveries today. When the $J/\psi$ was discovered in 1974, and the $\Upsilon(1S)$ in 1977, there was significant excitement due to what could be considered a BSM confirmation, in the sense that the standard theory was not settled on the necessary existence of charm quarks or bottom quarks until the discoveries were made.

In conclusion, it is not the nature of the SM feature confirmation itself that makes its discovery highly momentous in the progress of high-energy physics. Depending on circumstances a new hadronic resonance can signify a revolution (``November Revolution" of 1974~\cite{Barnett:2000} with $J/\psi$ discovery) or something far short of that ($\Sigma_b(6097)^\pm$ above).
Rather, the significance of a SM feature confirmation lies in the impact that the discovery has on BSM exclusion/confirmation immediately upon its discovery (such as Higgs boson and $J/\psi$) and the significant opportunities opened up for future BSM exclusion/confirmations through new experimental portals (such as through precision Higgs boson  and gravitational wave studies) recently made possible, which test BSM theories in ways previously inconceivable.

\xsection{European strategy update}

In this section we wish to add further justification to the articulation of various categories  of discovery (exclusion, confirmation, revolutionary) as presented above, and evidence that all these forms of discovery  are important to physicists. To accomplish the first we must show that particle physicists indeed use language that either mimics or evokes the categories described. To accomplish the latter, we must present evidence that physicists are willing to spend resources to bring about any of the various kinds of discovery. To a non-physicist the resources are monetary expenditures, but to the physicist it is their time, which is used to think and build, along with raising funds that enable them to think and build.

An excellent case study in which to investigate these questions is the European Strategy Updates. These occur in five-year intervals and are meant to set the agenda for European high-energy physics until the next update. This agenda is, at its core, a question of how resources are to be spent. This is enlightened, of course, by the science that physicists would like to accomplish. European budgets for particle physics, and CERN in particular, are relatively stable over time, and so the discussions are not about whether or not science should be done, but rather they are about exactly what science is to be done. This puts a healthy primary focus on what activities and pursuits scientists find most valuable, subject to reasonably well-known financial constraints.

We are presently in the midst of the 2018-2020 Update for European Strategy. A call for input to the process was made in early 2018 with a request that all written material to be submitted to a central repository by end of 2018. The submitted documents are then to be taken into account by the CERN council and other stakeholders in the planning of European high-energy physics projects. Deliberations and town hall meetings will occur during 2019 and a final report will be submitted in 2020. More details on the process can be found at its central website~\cite{EuropeanStrategy}.

\subsection{Framing the strategy update}

To obtain an initial understanding of the community's conceptualization of worthwhile discovery,  we look at the introductory document from the European Strategy committee, wherein they attempt to seed the discussion by saying (see ``about" tab linked at~\cite{EuropeanStrategy}), 
\begin{quote}
Understanding the properties of the Higgs boson (which was discovered at CERN just before the previous strategy update) remains a key focus of analysis at the LHC and future colliders, as are precision measurements of other SM parameters and searches for new physics beyond the SM. 
\end{quote}
In our language the first two activities (understanding Higgs properties and precision measurements of other SM parameters) has discovery value in locus confirmations or exclusions\footnote{For example, continued precision measurements of the top quark mass and the Higgs boson mass to determine if, under some simple assumptions, the universe is metastable~\cite{Degrassi:2012ry}.}, and also in the attempt to break the Standard Model, or in other words, make a SM falsification discovery. Increasing precision may cause increasing tension between observables that have to obey the correlative predictions of the Standard Model theory, and when that tension becomes too great the theory can no longer describe the data and is therefore falsified. It is not necessary that the precision measurements that falsified the SM be consistent with another theory in the theory canon for a discovery to be made, of course. All that was necessary to declare a discovery is that the SM was relegated to the dustbin. We should note that these two activities, as stated, do not guarantee discovery. They merely give the prospect for a SM falsification discovery. It would be a stretch and indignity to the usual usage of ``discovery" to call more and more precise measurements of the SM parameters to be an ``exclusion discovery."

One should pause to discuss again the asymmetric way in which the SM is viewed compared to other empirically adequate (BSM) theories, such as minimal supersymmetry with heavy-enough superpartners to have escaped detection, or the SMEFT, which has additional non-renormalizable operators in the theory beyond the SM's renormalizable ones. In the non-SM theories, experiments that exclude regions of parameter space are making exclusion discoveries, whereas excluding some regions of parameter space within the SM (such as narrowing the experimentally allowed top mass) is not considered here to be an exclusion discovery. This distinction reflects the sentiment of the community, which holds that the SM is a special reference theory that is the ``default correct" theory. Discoveries can only be made, if we are to use the word ``discovery" in a meaningful way, by either falsifying the SM, i.e.\ showing that no point in parameter space accommodates experimental measurements, or by reducing the parameter space of a non-SM theory, radically through confirmation or less radically by small exclusion improvements. 

The third activity listed in the European Strategy statement above is ``searches for new physics beyond the SM." This is a clear call for the pursuit of either revolutionary discovery of something not thought of by physicists before or through confirmation discovery with respect to an empirically adequate non-SM theory that resides presently within the theory canon. Revolutionary discovery would be great, but it is not possible to discuss it rigorously except to say that it would be interesting if something showed up that we have never thought about before. As we will see below that is the reason that with respect to searches for new physics beyond the SM, the discussion primarily centers on how a new experimental project may be able to make exclusion or confirmation discoveries within the theory canon.

\subsection{Discovery at colliders}

Many studies have been initiated and completed with the goal of contributing to the European strategy update. Contributions range from motivated small table-top experiments to next generation colliders on the energy frontier. In this subsection we will take a look at some of those contributions and ask in which ways their underlying conceptualization of discovery matches and differs from what has been described above. In particular, we look at the physics motivations for upgraded and new colliders. An excellent general argument for the utility of colliders is provided in~\cite{Giudice:2019zpk}. What follows below is an example analysis of specific discovery goals, and how they are characterized, at the high-luminosity and high-energy upgrades of the LHC (HL-LHC and HE-LHC, respectively).

\noindent \underline{\it BSM at HL/HE-LHC}

Perhaps the document that puts BSM physics most transparently at the center of its discussion is a report from working group 3 on ``Physics of the HL-LHC, and Perspectives at the HE-LHC" entitled ``Beyond the Standard Model Physics at the HL-LHC and HE-LHC"~\cite{CidVidal:2018eel}. The introduction strongly indicates that discoveries in the true sense of the word have already taken place at the LHC:
\begin{quote}
The lack of indications for the presence of NP [New Physics, i.e.\ BSM physics] so far may imply that either NP is not where we expect it, or that it is elusive. The first case should not be seen as a negative result. Indeed the theoretical and phenomenological arguments suggesting NP close to the electroweak (EW) scale are so compelling, that a null result should be considered itself as a great discovery.~\cite{CidVidal:2018eel}
\end{quote}
Here the authors have taken a slightly different view of discovery than has been advocated above. They implicitly require that  for a BSM exclusion to be labeled a discovery it requires that the new physics be ``expected" at the facility, which is a stronger requirement than our conceptualization of BSM exclusion discovery, which is perhaps best stated as an exclusion of parameter space that is ``not un-expected," which is an important distinction.

Now, it is not the intention in the above paragraph to declare that expectations for preferred parameter space regions  of BSM theories are not meaningful.  Rather, the intention is promote the label of discovery to BSM exclusions in parameter space regions that presently may not be considered {\it terra prima}, since assessments of coveted lands vs.\ non-coveted lands can change overnight for any number of reasons in physics, just as it can in geography (i.e., worthless arid lands may strike oil beneath). It should be noted that these views are not inconsistent with the report, but it is worthwhile explicitly making this point.

The BSM report continues 
\begin{quote}
A crucial ingredient to allow a comparison of proposed future machines is the assessment of our understanding of physics at the end of the HL-LHC program. Knowing which scenarios remain open at the end of the approved HL-LHC allows one to set standard benchmarks for all the interesting phenomena to study, that could be used to infer the potential of different future machines.~\cite{CidVidal:2018eel}
\end{quote}
In other words, it is the duty (``crucial ingredient") of anyone advocating for a new facility (``proposed future machines") to fully assess what prior experiment has done with respect to exclusions (what doesn't ``remain open") in the BSM theory canon (``scenarios"), and future facilities must have BSM exclusion capacity beyond that (``infer the potential").  This is fully in accord with our conceptualization of discoveries and the threshold of demonstrated scientific capabilities required of future machines.

The BSM report is almost exclusively devoted to the analysis and prospects of specific BSM scenarios where BSM confirmation/exclusion discoveries can be made. However, one section is devoted to ``signature based analyses", where it is introduced with
\begin{quote}
Several contributions that are constructed around experimental signatures rather than specific theoretical models are presented in this section. This includes analyses of dijets, diphotons, dibosons and ditops final state events at HL- and HE-LHC.
\end{quote}
This is an acknowledgement that some who are interested in making discoveries may  have ``signalistic" tendencies (see sec.~\ref{sec:signalism}), who are uncomfortable thinking directly about BSM theories and hope that revolutionary BSM discoveries can be made by analysing signatures (i.e., categories of events) that are manifestly incompatible with SM expectations. However, the irony of this section is that every one of these makes reference to BSM scenarios, illustrating our claim that one cannot escape BSM even if one tries. In the list below  the titles of each subsection are given along with the BSM physics explicitly invoked in the analysis, which of course goes well beyond just defining a signature:
\begin{itemize}
\item ``Coloured resonance signals at the HL- and HE-LHC" : introduces BSM diquarks into the spectrum.
\item ``Precision searches in dijets at the HL- and HE-LHC" : introduces a BSM ``new resonant state decaying to partons".
\item ``Dissecting heavy diphoton resonances at HL- and HE-LHC" : introduces ``a heavy resonance $X$, which decays via other new on-shell particles $n$ into multi- (i.e., three or more) photon final states."
\item ``Prospects for diboson resonances at the HL and HE-LHC" : introduces new ``resonances decaying to diboson ($WW$ or $WZ$, collectively called $V V$ where $V = W$ or $Z$ ) in the semileptonic channel where one $W$-boson decays leptonically and the other $W$ or $Z$-boson decays to quarks $(\ell\nu qq$ channel)."
\item ``Prospects for boosted object tagging with timing layers at HL-LHC" : introduces jets that obtain a mysterious BSM ``boost ... by $v=0.98$", which less mysteriously could be produced ``from $Z$ decay, where the $Z$ has been produced in the decay of a $1\tev$ diboson resonance."
\item ``High mass resonance searches at HE-LHC using hadronic final states" : introduces BSM ``new resonant states decaying to two highly boosted particles decaying hadronically."
\item ``On the power (spectrum) of HL/HE-LHC" : introduces BSM resonances that ``show up in Fourier space, after performing a Fourier transform on the relevant collider data."
\end{itemize}
One is tempted to conclude that ``signature based analyses" are really BSM exclusion/confirmation searches for  theories that are submissively recognized to not be prominent members the BSM theory canon, which it is perhaps hoped releases the practitioners from the responsibility of detailing them and defending them as legitimate theory targets of analysis. Although there might be an element of that at times, the  impetus is often to conduct an analysis that is relevant to many theories, with less close ties to any particular model. 

However, a  BSM-centered focus has a different perspective to ``signal based analyses." It is the BSM-specific theories that should be of interest primarily. BSM centered work recognizes that  a BSM theory leads to many signatures that experiment should cover, even signatures that we have not thought of before. A signal-based approach implicitly advocates the opposite direction of work: a particular signal can originate from many theories, even those theories you have never thought of before. Which approach is best? The BSM centered approach springs from solving {\it physics problems} (dark matter, baryogenesis, hierarchy, unification, flavor, etc.), whereas the signature-centered approach, when taken very seriously, springs from ``doing" something new with no other rationale. The signature-based approach is a slippery slope toward throwing vases off the Eiffel Tower to see what happens, as discussed in sec.~\ref{sec:signalism}. A detailed analysis of a BSM theory, on the other hand, puts pressure on defending the theory canon from whence it sprang, and it in no way impedes discovering other ``new physics scenarios" that are in the same signature class.

A different strategy to re-center the ``signal based" discussion of the report toward a BSM perspective, which would alter not just the presentation but also perhaps the work itself in places, would be to declare that these theories under discussion might be a bit odd, and that one cannot necessarily have much confidence perhaps that nature has chosen them, but they are empirically adequate, and each of them does satisfy at least some expectation criteria for new physics. Furthermore, they do give very different signatures compared to other theories under consideration, and perhaps would otherwise be missed if we did not take them into account. Thus, they should be admitted to the theory canon for further analysis. Now, this type of argument is indeed a reasonable argument  for admitting the theory to the BSM theory canon based on `diversity' of signatures predicted and theoretical humility~\cite{Wells:2018sus}. One should make that case explicitly. See if people will agree when all the information is laid out. They will agree if a decent case is made. There is no need to crouch behind pseudo-``signal based" categories.

\noindent \underline{\it SM at HL/HE-LHC}

The ``SM community" of researchers have also released a report: ``Standard Model Physics at the HL-LHC and HE-LHC"~\cite{Azzi:2019yne}. The main purpose of the document is to ``summarise the physics reach of the HL-LHC [and HE-LHC] in the realm of strong and electroweak interactions and top quark physics...."~\cite{Azzi:2019yne}. Furthermore, it is projected that a central task is to attempt to falsify the SM by the accrual of significant amounts of data to enable percent-level precision comparisons between theory and experiment. The reasons to do this are summed up in their introduction:
\begin{quote}
In addition, a considerable improvement is expected in precise measurements of properties of the Higgs boson, e.g. couplings measurements at the percent level, and of Standard Model (SM) production processes.... Anomalies in precision measurements in the SM sector can become significant when experimental measurements and theoretical predictions reach the percent level of precision, and when probing unprecedented energy scales in the multi-TeV regime. These anomalies could give insights to new physics effects from higher energy scales~\cite{Azzi:2019yne}.
\end{quote}

As we have discussed above, the desire to obtain higher and higher precision of observables is an obvious activity to stress-test and possibly falsify the SM. However, the SM parameters are already measured nearly as well as they will ever be measured at the LHC, even with higher luminosity. If we look at all the parameters of the SM, including the gauge couplings, Yukawa couplings, and the Higgs self coupling, there are only minor improvements to be had by HL-LHC. 

For example, the current top quark mass determination at ATLAS is $m_t=172.69\pm 0.48~\gev$~\cite{Aaboud:2018zbu} and at CMS $m_t=172.44\pm 0.48\gev$~\cite{Khachatryan:2015hba}. Preliminary CMS projects conclude that with $3\,{\rm ab}^{-1}$ the $14\tev$ HL-LHC can halve this uncertainty~\cite{Azzi:2019yne}. Likewise, the $W$ mass currently has an uncertainty of $12\,{\rm MeV}$~\cite{Tanabashi:2018}. Projections of uncertainty projections for the $W$ mass at HL-LHC are approximately 50\% improvement. Although this is clearly progress, it is not possible to assess what level of progress or significance this improvement has without reference to BSM physics. 

The significance of precision $W$ boson measurement, electroweak vector boson production rates and differential cross-sections at  higher energies are especially dependent on BSM context. Within a pure SM mindset, such measurements {have no special motivation or value} if they do not increase the precision of any SM parameter above what was done long ago at LEP. To know what measuring $W^+W^-$ scattering at the highest possible energy is more valuable than improved measurements of the $\Lambda_b$ mass  --- neither one of which improves knowledge of the SM itself at all --- comes only from recognizing that BSM theories are more likely to insert themselves and disrupt the SM expectations of the high-energy $\sigma(W^+W^-)$ measurement than the $m_{\Lambda_b}$ measurement.

It is for these reasons discussed above that the ``SM at LHC" document contains so much discussion on BSM physics. There is discussion of four-fermion operators, pseudoscalar colour-octets, exotic top quark dipole moments, exotic non-SM quartic gauge couplings, exotic diphoton and dilepton resonances, etc. The discussion signifies and implicitly declares that SM analyses (experimental and theoretical) can have no interesting justification and no profound value  without a  BSM context.

\subsection{Discovery beyond colliders}

In addition to projects that involve colliders, there are many interesting projects that impact the high-energy physics frontier that do not involve colliding particles at the highest possible energies. This is abundantly clear in the neutrino physics programs around the world, which focus on high-precision measurements by detectors sensitive to various species of neutrinos at various energies. Astrophysical measurements aimed at detecting non-standard deviations in cosmic microwave radiation, cosmic strings, axions and dark matter are also part of this scientific endeavor.

One recent document submitted for consideration during the European Strategy Update is the ``Summary Report of Physics Beyond Colliders at CERN"~\cite{Alemany:2019vsk}. Many projects and ideas are presented. An implicit recurring theme is that for a project to be justified it must have BSM  exclusion/confirmation discovery capability that extends beyond any other experiment of the past, or of other experiments (including colliders) approved on the horizon. 

The focus, therefore, is on BSM exclusion/confirmation of theories that are very weakly coupled to the SM (``dark sectors") and especially those that are of small mass and thus buried in the background of the more traditional collider experiments. New, lower-energy experiments are needed with specialized capacity to make these kinds of BSM discoveries. Examples of BSM targets are ``are dark matter, messenger particles to dark matter, explanations of the $(g - 2)_\mu$ anomaly, the proton radius anomaly, stellar cooling anomalies and many more"~\cite{Alemany:2019vsk}.

Focus on BSM discoveries has tangible implications. Instead of randomly selecting what features a new detector might have, which a revolutionary or SM falsification orientation would necessarily entail, a BSM-centered approach makes deliberate and justified choices in order to make BSM exclusion/confirmation. An example is the SHiP detector, which introduces many specialized features for BSM purposes. For example, the report details some key features of SHiP introduced to maximize sensitivity to particular BSM scenarios:
\begin{quote}
In addition to the mainstream spectrometer, SHiP is planned to be equipped with a high precision emulsion spectrometer located immediately upstream of the decay vessel. This subdetector will increase the discovery reach by providing sensitivity to re-interactions of long lived particles produced in the dump, and will collect a first high statistics sample of $\tau$-neutrino interactions to test lepton universality~\cite{Alemany:2019vsk}.
\end{quote}
Such specific design choices with BSM tests firmly in mind are detailed for all the experimental suggestions given. 

Furthermore, it is emphasized in this report that the new experimental ideas are complementary to other experiments. They can make BSM exclusion/confirmation discoveries that no other experiment can. This is graphically emphasized in many figures of the report~\cite{Alemany:2019vsk}, including Fig.~17 in the case of projected sensitivities to dark scalars, where different experiments have coverage of different regions of parameter space. Without a BSM context there can be no such considerations. One could just as easily claim that every experiment ever contemplated is unique (which they are in some way) and thus has just as much claim to become realized as any other experiment. There can be little rational basis for making decisions about any experiment without an explicitly invoked BSM context.

\subsection{BSM theory}

A straightforward observation from reading reports that hope to influence the European Strategy Update is that BSM exclusion/confirmation discoveries are still central pursuits in the experimental realm. There is recognition, albeit implicitly  at times, that any justification of new experimental project necessarily involves understanding and demonstrating its BSM exclusion/confirmation capabilities. There can be little rational basis for any such decision without BSM.

Despite the centrality of BSM physics, there  is very little discussion on the need to invest in a balanced BSM theory program. The BSM theory community is not a laboratory with a director general. It is not a single large collaboration with hundreds or thousands of members with a spokesperson. It is a collection of (near) solitary pursuits that are centrally vital to the progress of high-energy physics. A sure way to diminish high-energy physics despite the presence of sufficient financial resources is to allow BSM theory to stagnate and diminish by reducing its priority. It is easy to do since there are no institutionally constructed leaders to keep its value prominently reminded, as there is in large experimental collaborations. One must keep in mind, however, that the stagnation and subsequent diminishment of BSM theory means the diminishment of high-energy experiment (and vice-versa of course), and with the loss of BSM theory comes the loss of defensible rationales for transformative experiment.

\xsection{When does discovery end?}

Since the apotheosis of science hundreds of years there have been conjectures, worries, ``proofs", and declarations that discovery has ended or will end soon~\cite{Wells:2016a}. It is interesting to consider what conditions must be met to feel confident that scientific discovery has come to an end, using the language of discovery developed here. In this conceptualization,  proof that scientific discovery ends is fulfilled when all three ``propositions of discovery cessation" are proven to be true.

\vspace{0.2cm}
{\bf Propositions of discovery cessation} \vspace{-0.4cm}
\begin{enumerate}
\item There is no prospect for exclusion/confirmation discovery.  
\item There is no prospect for SM falsification.
\item There is no prospect for revolutionary discovery.
\end{enumerate}

Proving propositions 2 and 3 appear to be formally impossible. Regarding proposition 3 in particular, we have argued above that seeking revolutionary discoveries outside the context of BSM exclusion is mystical in nature, and the arrival of revolutionary discoveries is rare and can never be guaranteed. There is no theory for the expected rate of revolutionary discoveries, and thus there can be no assessment that we will fail to have one again in the future. 

Proposition 1 has been argued above as the key proposition for guaranteeing discovery which directly counters any cessation claims. Thus, a practical discussion of whether one is at the end of discovery can be had by assessing the status of BSM theories within the theory canon. In that spirit, cessation proposition 1 could be true for any number of reasons, including these three ``conditions":

\vfill\eject

{\bf Conditions of discovery curtailment}\vspace{-0.4cm}
\begin{itemize}
\item[a)] The theory canon becomes empty of any viable or motivated BSM theory, including BSM-motivated loci of SM parameter space. 
\item[b)] Or, if there are BSM theories in the theory canon, there nevertheless is no idea for how to turn reasonable resources of time, people, and money into an experiment that can make exclusion/confirmation discoveries within the canon. 
\item[c)] Or, if there exist BSM theories in the theory canon and there exist reasonable experimental ideas to make exclusion/confirmation discoveries, there nevertheless exist insufficient resources to pursue them, such as no willingness by governments to financially invest in discovery, or no willingness of scientists to invest sufficient time to pursue the future discoveries. 
\end{itemize}

It is abundantly clear that we are far from reaching the discovery curtailment conditions a) and b). As the technical design reports of the European Strategy process confirm, there is a multitude of interesting BSM theories within the theory canon that are empirically adequate and which can be excluded/confirmed by future proposed experiments\footnote{And it should be emphasized that the standard for interest in BSM theories is not that they are guaranteed to be found at the next future experiment if they exist, but rather that they purport to solve a problem or some other claim to expectation, and that they have a reasonable, but not necessarily guaranteed, prospect for their effects to be discerned.}. The high-energy physics endeavor is also far from reaching curtailment condition c). However, it should be noted that condition c) is frequently the highest risk when it comes to pursuing discovery in science, and it is one that needs as much diligence staving off as the a) and b) conditions, even though a) and b) are more enjoyable to pursue by the scientists.

In short, there is no justifiable claim that the end of discovery is now or nigh, and there is no present sign of asymptoting to any of the discovery curtailment conditions. Nevertheless, the diminishing of effort toward constructing and maintaining a vibrant BSM theory canon would work to activate curtailment condition a), the diminishing of investment in devising experimental ideas and methods (including detector and accelerator development) would work to activate curtailment condition b), and the abandonment of proposing and lobbying for future state-of-the-art facilities that guarantee exclusion/confirmation discovery would work to activate curtailment condition c). Staving off the end of discovery requires effort on many complementary fronts as a community.

\xsection{Summary}

Below is a summary of the main arguments contained in this essay, some of which are rather obvious but need to be articulated for coherence, while others are the result of more fully developed provocations above. 

\begin{itemize}
\item No discovery happens without the existence of a theory canon populated by the SM and empirically adequate BSM theories.
\item The three main direct discovery activities of high-energy physics are theory canon building (i.e., model building), experiment building (i.e., creative design, construction and execution), and analysis, which connects canon theories to experimental possibilities  and experimental results to theory expectations.
\item Among the three broad types of discovery --- confirmation, exclusion, and revolutionary --- only exclusion discovery can be guaranteed.
\item A BSM exclusion discovery is a significant experimental feat since by definition it requires exclusion beyond what all prior experiment has been able to accomplish.
\item BSM confirmation discoveries can only happen by experiments that had guaranteed BSM exclusion capability at the outset.
\item BSM falsification claims for decoupling theories (i.e., those whose parameter spaces allow observable predictions arbitrarily close to those of the SM) often require the application of strong non-empirical constraints to their parameter spaces, such as simplicity and lack of finetuning, and thus are of controversial significance and are inherently polemical from the strict empirical perspective.
\item BSM falsifications for theories that are defined with no strong non-empirical constraints on their parameter spaces are secure and non-controversial falsification claims, since the parameter space boundaries are not up for controversial non-empirical re-assessments.
\item Approval for any future high-investment research facility should be reserved for those with guaranteed discovery, or in other words, with demonstrated capacity to make BSM exclusion discoveries with respect to  the theory canon.
\item The main methodological rival to exploration centered on BSM exclusion/confirmation is signalism, which aims through ``signal-based analysis" and ``model independent searches" to achieve SM falsification or revolutionary discoveries without any reference to BSM theories.
\item Signalism's approach to pursue only revolutionary discoveries without reference to BSM theories leads to inscrutable mystical exertions that have no rational claim to utility, while at the same time poses unwelcome risks to BSM confirmation/exclusion discoveries and SM falsification.
\item Signalism's closely related other approach, to pursue only SM falsification without reference to any BSM theories, poses risks to every type discovery, including SM falsification, while focus on BSM exclusion/confirmation discovery heightens prospects for SM falsification.
\item The SM is just one empirically adequate theory among many, and thus the asymmetric worship of the SM and disdain for BSM theories among many signalism-oriented physicists is an ironical manifestation of a rigid extra-empirical philosophy of theory choice. 
\item The prospects for truly revolutionary discoveries are unlikely to be adversely affected by focus on BSM exclusion/confirmation compared to any other approach hoping to maximize revolutionary discovery's chances.
\item Excitement for some recent SM feature confirmations (e.g., Higgs boson and gravity waves) versus others (e.g., new hadronic resonance) is not to be understood by intrinsic SM-based worth criteria, but rather is due to the BSM exclusion/confirmations that are made possible presently and into the future by the result, which further reveals how central BSM is to excitement, interest and intuited notions of progress.
\item Discovery hopes cease when {\it all} of the following three conditions are shown to be true: \\
1) there is no prospect for BSM exclusion/confirmation discoveries; \\
2) there is no prospect for SM falsification discovery; {and}, \\
3) there is no prospect for revolutionary discoveries. \\
We are presently far from justifying any of these three propositions of discovery cessation.
\item Discovery can be curtailed when {\it any} of the following three conditions is present: \\
1) the BSM theory canon is allowed to languish; \\
2) experimental ideas to confirm/exclude the BSM canon diminish; or,  \\
3) governmental/institutional support diminishes.\\
All three require continual attention to maintain healthy discovery in high-energy physics.
\end{itemize}

High-energy physics is rather unique among the sciences, as it is by construction a field of inquiry that pursues a frontier that is by all practical definitions infinite, and thus promises mystery and anticipation for as far as the mind can see.  As such, its primary goal is to make discoveries akin to sea-faring explorers of the past --- a journey complete with financiers, officers, subalterns, visionaries, and mutineers. As pointed out above, discoveries are only possible through the passage of BSM exclusions, just as Ponce de Le\'on's discovery of Florida was made possible only through passage of open seas. High-energy physics has expertly charted the high seas for some time, noting  the fascinating islands of recent confirmation discoveries that include the charm quark, the top quark, massive neutrinos, and the Higgs boson. We sail further. Maybe there is a near-by continent up ahead. Let's see.

\medskip\noindent
{\it Acknowledgments:}  The author wishes to thank R.\ Akhoury, S.P.\ Martin and A.\ Pierce for helpful conversations on these issues. Support provided in part by the DOE (DE-SC0007859) and by the Alexander von Humboldt Foundation when this work was initiated.

\end{document}